\documentclass[aps,prc,showpacs,showkeys,nofootinbib]{revtex4}

\usepackage{graphicx}
\usepackage{dcolumn}
\usepackage{amsthm}
\usepackage{bm}

\newcommand{\definition}{\textit}

\begin{document}

\title{Quantum corrections to the Relativistic Mean-Field theory}


\author{Sergei~P.~Maydanyuk$^{(1,2)}$}%
\email{maidan@kinr.kiev.ua}%
\author{Peng-Ming~Zhang$^{1}$}%
\email{zhpm@impcas.ac.cn} %
\author{Ahmed~Bakry$^{1}$}%
\email{abakry@impcas.ac.cn}%
\affiliation{$(1)$Institute of Modern Physics, Chinese Academy of Sciences, Lanzhou, 730000, China}
\affiliation{$(2)$Institute for Nuclear Research, National Academy of Sciences of Ukraine, Kiev, 03680, Ukraine}

\date{\small\today}


\begin{abstract}
%
In this paper, we have compared the RMF theory and the model of deformed oscillator shells (DOS) in description of the quantum properties of the bound states of the spherically symmetric light nuclei.
We have obtained an explicit analytical relation between differential equations for the RMF theory and DOS model, which determine wave functions for nucleons.
On such a basis we perform analysis of correspondence of quantum properties of nuclei.
Conclusions are the following.
%
(1) On the basis of the relations above, we found the potential $V_{RMF}$ of the RMF theory for nucleons which has the wave functions $f$ and $g$ for nucleons with joint part $h$ coincident exactly analytically with the nucleon wave function of DOS model with potential $V_{\rm shell}$.
This fact confirms that it is possible to describe quantum properties of nucleus in framework of the RMF theory exactly the same as in DOS model.
However, a difference between $V_{RMF}$ and $V_{\rm shell}$ is essential for any nucleus.
%
%
(2) The nucleon wave functions and densities obtained by the DOS and RMF theories are essentially different.
The nucleon densities of the RMF theory contradict to knowledge about distribution of the proton and neutron densities inside the nuclei obtained from experimental data.
This indicates that $g$ and $f$ have no sense of the wave functions of quantum physics.
But, $h$ provides proper description of quantum properties of nucleons inside the nucleus.
%
%
(3) We calculate meson function $w^{0}$ (and corresponding potential $V_{w}$ in framework of RMF theory) based on the found nucleon density above.
%
%
(4) $f$ and $g$ are not solutions of Dirac equation with $V_{w}$.
If the meson theory describes quantum properties of nucleus well, then a difference between $V_{w}$ and $V_{RMF}$ should be as small as possible.
We introduce new quantum corrections 
characterizing difference between these potentials.
We find that
(a) The function $w^{0}$ should be reinforced strongly,
(b) The corrections are necessary to describe the quantum properties of the nuclei. 
\end{abstract}


\maketitle

\section{Introduction
\label{sec.introduction}}

Nuclear interactions are one of the most fundamental subject in physics, which attracted scientists for a very long time and a lot of efforts were put to their understanding. Here, many theories have been developed. In this paper we put our interest to two different theories.

The first one is the Relativistic mean-field (RMF) theory (see Refs.~\cite{Ring.1990.AP.v198,Ring.1996.PPNP,
Ring_Afanasjev.1997.PPNP,Ring.2001.PPNP,Ring_Serra_Rummel.2001.PPNP,
Ring.2007.PPNP,PenaArteaga_Ring.2007.PPNP,Ring.2011.PPNP,Ring.2005.PRep} and reference therein)
which is based on the relativistic formulation of unified lagrangian, combining nucleons of the studied nuclear system, while nuclear interactions are described via exchange of mesons of different types.
This theory, initially born in the 1970s from works of John Dirk Walecka on quantum hadrodynamics, from end of 1980s has been intensively investigated by Piter~Ring and coworkers and then many other researchers.
It seems this theory is one of the most actively developed and successful in nuclear physics, as
it now well describes all known properties of nuclear structure
(for example, see
general properties~\cite{%
Ring.1997.PRC.v55.p540,
Ring.1999.ADNDT.v71,
Ring.1974.NPA.v235,
Boguta.1977.NPA.v292
},
deformations and asymmetries~\cite{%
Shan-Gui_Zhou.2012.PRC.v86,
Bhuyan.2015.PRC}, 
superdeformations~\cite{%
Ring.1997.PRC.v55.p540,
Ring.1990.NPA.v511,Ring.1996.NPA.v608,Bonche.1996.NPA},
magnetic multipole moments and transition probabilities~\cite{Ring.1973.NPA.v209}, 
magnetic rotation bands~\cite{Ring.2011.PLB.v699}, 
collective rotational motion in deformed nuclei~\cite{Ring.1989.NPA.v493}, 
shell structure~\cite{Bender.1999.PRC}, 
pseudospin symmetry for bound states~\cite{Shan-Gui_Zhou.2013.PRC.v88}, 
anomaly in the charge radii~\cite{Ring.1993.PLB}, 
phase transitions in nuclear matter~\cite{Muller.1995.PRC}, 
etc.),
is applied for analysis of nuclear reactions
(for example, see for
fission~\cite{Shan-Gui_Zhou.2012.PRCrapid.v85,Shan-Gui_Zhou.2014.PRC.v89,Shan-Gui_Zhou.2015.PRC.v91,Shan-Gui_Zhou.2016.PRC.v93}%
),
is used in different tasks of
nuclear astrophysics~\cite{%
Danielewicz.2002.Science,
Chin.1977.AP,
Shen.1998.NPA,
Bhuyan.2014.JPG}, 
exotic nuclei~\cite{Ring.1993.PLB.v312,Ring.2005.PRep},
hypernuclei~\cite{Shan-Gui_Zhou.2011.PRC.v84},
etc.
In frameworks of such a theory, nucleons obey the Dirac equation, while the mesons obey the Klein-Gordon equations (i.e. equations of motion).
Essential efforts are put to understanding of
nuclear matter, symmetry energy and binding energy of nuclei in the region from the lightest to heavy, and even
superheavy~\cite{Sugahara.1994.NPA.v579,
Ring.2003.PRC.v68
}.

The second theory is a microscopic nuclear theory of nucleus (MN theory) which is unified theory
oriented on the fully quantum basis in description of properties of the nuclear structure, scattering and reactions
\cite{Wildermuth.1977.book,Tang.1978.PR,Tang.1981.lectures}.
It is very effective in detailed study and description of clustering phenomena in nuclei.
The resultant formalism is now generally known as the Generator coordinate method~\cite{Horiuchi.1977.PTPS,Filippov.1984.PEPAN,Filippov.1985.PEPAN}
introduced by Hill, Griffin and Wheeler~\cite{Hill.1953.PR,Griffin.1957.PR}.
However, it should be mentioned, that this method is equivalent to the Resonating group method~%
\cite{Tang.1978.PR}
proposed by Wheeler in Refs.~\cite{Wheeler.1937.PR.p1083,Wheeler.1937.PR.p1107}.
%
%
Since 1970-s, a model of the deformed oscillator shells (DOS) as a part of this theory was successfully developed for study of bound states of nuclear systems
\cite{Steshenko.1971.SJNP}
(note an Algebraic model of the resonating group
method~\cite{Vasilevsky.1997.PRA,Vasilevsky.2001.PRC.part1,Vasilevsky.2001.PRC.part2,Vasilevsky.2001.PRC.p064604,
Vasilevsky.2005.PRC,Vasilevsky.2010.PRC,Vasilevsky.2012.PRC}
which is a unified approach to describe bound and continuous states of nuclear systems,
%
a microscopic continuum-discretized coupled-channels reaction approach~\cite{Descouvemont.2014.PRC,Descouvemont.2016.PRC.v90,Descouvemont.2016.PRC.v93}
oriented on the microscopic description of reactions,
see also reference therein).
Such a model represents the nuclear many-particle wave functions through their expansions in harmonic oscillator eigenstates
(with application of Pauli principle),
where nucleon-nucleon potentials are parameterized from data of the experimental nucleon-nucleon scattering.
In frameworks of such a model, the light nuclei starting from $^{4}{\rm He}$ were investigated, with a good agreement with known experimental information for them.
This model clearly well describes physics of the light nuclei (binding energy, sizes, deformations, quadruple momenta, etc.)
and, so, we shall focus on it in study of bound states of these nuclei in this paper.
It is important that this model forms a clear accurate quantum basis for more complicated unified microscopic nuclear theory
in description of bound states of the light nuclei.

At a brief comparison of these two theories one can find, that wave function of nucleons in the RMF theory is solution of Dirac equation (with interacting meson potential), while a corresponding nucleon wave function in the MN theory is solution of the Schr\"{o}dinger equation.
This fact provides a principal difference between properties and behavior of these wave functions of two theories:
these two wave functions are never identical at any arbitrary interacting potentials, in principle.%
\footnote{We prove this statement by Eqs.~(\ref{eq.5.1.9}), see below in the text.
Exception is for free particle moving in vacuum only.}
By such a reason, one can suppose that this difference should be shown both in different description of nuclear structure (bound states), and moreover in nuclear reactions (unbound states, i.e. scattering, decay, capture states, etc.).
In particular, it becomes clear from such a logic that nucleon densities in nuclei are always different in framework of two theories.
As the wave function is key characteristic of quantum properties of the studied object in quantum mechanics,
one can conclude that these two theories cannot provide the same quantum description of the nuclei and nuclear reactions.%
\footnote{One can remind that Dirac equation was written for description of free point-like fermion in vacuum in order to include relativistic relations between space and time, while the Schr\"{o}dinger equation was devote to the most accurate description of the quantum properties of the studied (spinless) system (which can be complicated) in the interacting potential.}

So, a main aim of this paper is to clarify and resolve this puzzle.
We analyze and compare these two theories in their quantum description of the nuclear properties and interactions.
In order to find key idea and analyze it as clearly as possible, we choose the light nucleus $^{4}{\rm He}$ for our analysis and calculations.
As this nucleus is spherically symmetric, we can apply the spherically symmetric approximation that reduces essentially number of equations.
It turns out that one can find an explicit analytical relation between differential equations for the RMF theory and DOS model, which determine wave functions for nucleons.
These relation is powerful tools for performing a comparative quantum analysis of these two theories.
As a mathematical basis of the RMF theory for analysis in this paper,
we use equations of motions as Eqs.~(36)--(40) in Ref.~\cite{Ring.1996.PPNP},
densities as Eqs.~(41)--(44) in Ref.~\cite{Ring.1996.PPNP},
potentials as Eqs.~(45)--(46) in Ref.~\cite{Ring.1996.PPNP}.

This paper is organized by the following.
In Sec.~\ref{sec.2.2} and~\ref{sec.3} we obtain an explicit analytical relation between differential equations for the RMF theory and DOS model, which determine wave functions for nucleons.
Here, we present a new approach for determination of a transformed new potential and part of the Dirac wave function corresponding to the potential and wave function of the shell model.
A comparative analysis of the potential of the RMF theory found by such a way with a corresponding potential of the shell model is given in Sec.~\ref{sec.4}. We start practical calculations and analysis on the example of the $^{4}{\rm He}$ nucleus, which is used in demonstrations of
all next found characteristics.
The wave function and corresponding density for nucleons are defined and calculated in Sec.~\ref{sec.5}. Here, we add an series expansion approach for determination of these characteristics,
which is effective in next determination of the meson potential and quantum corrections.
The meson potential based on the nucleon density found above is obtained in Sec.~\ref{sec.6}, the quantum corrections are defined and calculated in Sec.~\ref{sec.7}.
In Sec.~\ref{sec.conclusions} we summarize the paper.
Some needed details of calculations are added in Appendixes~\ref{sec.app.1}--~\ref{sec.app.5}.

\section{Spherically symmetric consideration
\label{sec.2.2}}

We start from equation of motion of the RMF theory for nucleon with number $i$ defined in Eq.~(36) in Ref.~\cite{Ring.1996.PPNP}
(in this section we remove the index of numeration of nucleons):
\begin{equation}
\begin{array}{lcl}
  \bigl\{- i\, \alpha \nabla\, + \beta\, (m_{0} + S (\mathbf{r})) + V (\mathbf{r})) \bigr\}\, \psi(\mathbf{r}) = \varepsilon\, \psi(\mathbf{r}), \\
\end{array}
\label{eq.2.2.1}
\end{equation}
where $m$ is mass of nucleon,
$V(\mathbf{r})$ and $S(\mathbf{r})$ are potentials defined in Eqs.~(45) and (46) in Ref.~\cite{Ring.1996.PPNP}.
If $V(r)$ and $S(r)$ are radial functions,
we look for a solution for wave function in form (see Sect.~1.5.1, page~48 in Ref.~\cite{Ahiezer.1981}):
\begin{equation}
\begin{array}{lcl}
  \psi_{\varepsilon jlM}(\mathbf{r}) =
  \left(
  \begin{array}{lcl}
    \varphi_{\varepsilon jlM} (\mathbf{r}) \\
    \chi_{\varepsilon jlM} (\mathbf{r})
  \end{array}
  \right),
\end{array}
\label{eq.2.2.2}
\end{equation}
where
\begin{equation}
\begin{array}{lcl}
  \varphi_{\varepsilon jlM} (\mathbf{r}) =
   g\,(r)\; \Omega_{jlM} (\mathbf{n}), &
  \chi_{\varepsilon jlM} (\mathbf{r}) =
   i\,f\,(r)\; \Omega_{jl^{\prime}M} (\mathbf{n}),
\end{array}
\label{eq.2.2.3}
\end{equation}
and $\Omega_{jlM} (\mathbf{n})$ is \definition{spherical harmonic spinor} (called also as \definition{spherical harmonic spinor function}), $l^{\prime}=2j-l$, $\mathbf{n} = \mathbf{r}/r$, $r=|\mathbf{r}|$, $g\,(r)$ and $f\,(r)$ are radial functions.
According to Ref.~\cite{Ahiezer.1981}, $\varepsilon$ is eigenvalue for Hamiltonian (but it is not energy for particle), which can be positive and negative.
At $\varepsilon>0$ equations describe particle, at $\varepsilon<0$ --- antiparticle.
Energy is defined as $E = |\varepsilon|$, i.~e. it is positive for particles and antiparticles.

In the case when scalar potential $S (r) = S_{0}$ is constant,
Eq.~(\ref{eq.2.2.1}) can be transformed explicitly to the differential equation of the second order
(Schr\"{o}dinger type equation) as (see Appendix~\ref{sec.app.1}):
\begin{equation}
  -\,h^{\prime\prime} + \bar{V}(r)\, h = 0,
\label{eq.2.2.4}
\end{equation}
where
$h(r)$ is a new unknown wave function and
a new potential is introduced as
\begin{equation}
\begin{array}{cc}
  \bar{V}(r) =
  \displaystyle\frac{V^{\prime\prime}}{2\, (\varepsilon + m - V)} +
  \displaystyle\frac{3\,(V^{\prime})^{2}}{4\, (\varepsilon + m - V)^{2}} -
  \displaystyle\frac{V^{\prime}}{\varepsilon + m - V(r)}\, \displaystyle\frac{\chi}{r}\, -
  \bigl[ \varepsilon - V(r) \bigr]^{2} + m^{2} +
  \displaystyle\frac{\chi (1 + \chi)}{r^{2}},
\end{array}
\label{eq.2.2.5}
\end{equation}
and
$m = m_{0} + S_{0}$,
$\chi = l(l+1) - j(j+1) - 1/4$ (see~(1.4.6), page~44 in~\cite{Ahiezer.1981}).
The unknown new wave function $h(r)$ and its derivative $h^{\prime}(r)$ should be obtained via numeric solution of Eq.~(\ref{eq.2.2.4}) with the potential $\bar{V}(r)$.
Then, the old wave function is obtained as (see Appendix~\ref{sec.app.1})
\begin{equation}
\begin{array}{lcl}
  g\,(r) & = & \displaystyle\frac{n_{0}\; \sqrt{|\varepsilon + m - V(r)|}}{r} \cdot h\,(r), \\
  f\,(r) & = &
    {\rm sign}\, \bigl[\varepsilon + m - V(r) \bigr]\;
    \displaystyle\frac{n_{0}}{r\: \sqrt{|\varepsilon + m - V(r)|}}\;
    \biggl\{
      \Bigl[
        \displaystyle\frac{-V^{\prime}(r)}{2\: (\varepsilon + m - V(r))} +
        \displaystyle\frac{\chi}{r}\;
      \Bigr]\; h\,(r) +
      h^{\prime}\,(r)
    \biggr\},
\end{array}
\label{eq.2.2.6}
\end{equation}
where
$n_{0}$ is free arbitrary constant and
\begin{equation}
  {\rm sign}\, \bigl[x \bigr] =
  \left\{
  \begin{array}{ll}
    1,  & \mbox{\rm at } x > 0, \\
    -1, & \mbox{\rm at } x < 0.
  \end{array}
  \right.
\label{eq.2.2.8}
\end{equation}
Also we have a property
${\rm sign}\, \bigl[- x \bigr] = -\, {\rm sign}\, \bigl[x \bigr]$.
%


We rewrite the solutions above via dimensionless variable.
In this connection, it is convenient to introduce, instead of the coordinate $r$, a new \emph{dimensionless} variable $\xi_{r}$ by the relation
\begin{equation}
  \xi_{r} = \displaystyle\frac{r}{a_{r}},
\label{eq.2.3.1}
\end{equation}
where
$a_{r}$ is a new coefficient defined from the shell model for the chosen studied nucleus.
After transformations Eq.~(\ref{eq.2.2.4}) is transformed to
\begin{equation}
  -\,\displaystyle\frac{d^{2} h}{d\xi_{r}^{2}} + \bar{V}_{\xi}(\xi_{r})\, h = 0,
\label{eq.2.3.2}
\end{equation}
where
\begin{equation}
\begin{array}{cc}
  \bar{V}_{\xi}(\xi_{r}) =
  \displaystyle\frac{V^{\prime\prime}}{2\, (\varepsilon + m - V)} +
  \displaystyle\frac{3\,(V^{\prime})^{2}}{4\, (\varepsilon + m - V)^{2}} -
  \displaystyle\frac{V^{\prime}}{\varepsilon + m - V}\, \displaystyle\frac{\chi}{\xi_{r}} -
  a_{r}^{2} \bigl[ \varepsilon - V \bigr]^{2} + a_{r}^{2} m^{2} +
  \displaystyle\frac{\chi (1 + \chi)}{\xi_{r}^{2}},
\end{array}
\label{eq.2.3.3}
\end{equation}
the wave radial functions $g(\xi_{r})$ and $f(\xi_{r})$ have the following solutions:
\begin{equation}
\begin{array}{lc}
  g\,(\xi_{r}) = \displaystyle\frac{n_{\xi 0}\; \sqrt{|\varepsilon + m - V(\xi_{r})|}}{\xi_{r}} \cdot h\,(\xi_{r}), \\

  f\,(\xi_{r}) =
    {\rm sign}\, \bigl[\varepsilon + m - V(\xi_{r}) \bigr]\;
    \displaystyle\frac{n_{\xi 0}}{a_{r} \xi_{r}\: \sqrt{|\varepsilon + m - V(\xi_{\xi_{r}})|}}\;
    \biggl\{
      \Bigl[
        \displaystyle\frac{-V^{\prime}(\xi_{r})}{2\: (\varepsilon + m - V(\xi_{r}))} +
        \displaystyle\frac{\chi}{\xi_{r}}\;
      \Bigr]\; h\,(\xi_{r}) +
      h^{\prime}\,(\xi_{r})
    \biggr\}
\end{array}
\label{eq.2.3.4}
\end{equation}
and $n_{\xi 0}$ is a new arbitrary constant of integration.

\section{Determination of potential $V$ on the basis of the given $\bar{V}$
\label{sec.3}}

We shall rewrite the unknown potential $V(r)$ on the basis of the known potential $\bar{V}(r)$.
We introduce the following change of potential:
\begin{equation}
   V_{1}(r) = V(r) - \varepsilon - m
\label{eq.3.1.2}
\end{equation}
and from Eq.~(\ref{eq.2.2.5}) we obtain
\begin{equation}
\begin{array}{cc}
  \bar{V}(r) =
  -\, \displaystyle\frac{V_{1}^{\prime\prime}}{2\, V_{1}} +
  \displaystyle\frac{3\,(V_{1}^{\prime})^{2}}{4\, V_{1}^{2}} +
  \displaystyle\frac{V_{1}^{\prime}}{V_{1}}\, \displaystyle\frac{\chi}{r}\, -
  \bigl[ V_{1} + m \bigr]^{2} + m^{2} +
  \displaystyle\frac{\chi (1 + \chi)}{r^{2}}.
\end{array}
\label{eq.3.1.3}
\end{equation}
In case of the spherically symmetric nuclei in the ground state we have:
\begin{equation}
\begin{array}{ccc}
  l = 0, & j = \displaystyle\frac{1}{2}, & \chi = -1
\end{array}
\label{eq.3.1.4}
\end{equation}
and
%
%
%
\begin{equation}
\begin{array}{cc}
  4\, \bar{V}(r)  V_{1}^{2} =
  - 2\, V_{1} V_{1}^{\prime\prime} +
  3\,(V_{1}^{\prime})^{2} -
  \displaystyle\frac{4}{r}\, V_{1} V_{1}^{\prime}\, -
  4\, V_{1}^{4} - 8\, V_{1}^{3}\, m.
\end{array}
\label{eq.3.1.6}
\end{equation}
The next task is to find the unknown $V_{1}$ on the basis of the given $\bar{V}$.

%
%
%

\subsection{Solutions in form of series
\label{sec.3.2}}

We shall analyze a possibility of existence of solution of equation (\ref{eq.3.1.6}) in form of series:
\begin{equation}
\begin{array}{lcl}
\vspace{1mm}
  V_{1}(r) & = & f_{0} + \displaystyle\sum\limits_{n=1}^{N} f_{n} r^{n}, \\
\vspace{1mm}
  V_{1}^{\prime}(r) & = &
    f_{1} + \displaystyle\sum\limits_{n=1} (n+1) f_{n+1} r^{n}, \\
  V_{1}^{\prime\prime}(r) & = &
    2f_{2} +
    \displaystyle\sum\limits_{n=1} (n+1)(n+2) f_{n+2} r^{n}.
\end{array}
\label{eq.3.2.1}
\end{equation}
We consider potential in the following form:
\begin{equation}
  \bar{V} (r) = a_{0} + a_{1} r + a_{2} r^{2},
\label{eq.3.2.2}
\end{equation}
where $a_{0}$, $a_{1}$ and $a_{3}$ are parameters, defined by the microscopic shell model of nucleus.
Taking into account the initial equation~(\ref{eq.3.1.6}), we obtain equation for determination of the unknown amplitudes $f_{n}$.
Solving this equation (see App.~\ref{sec.app.3}, for details), we obtain such amplitudes
(at $n \ge 2$, also we shall use $f_{-1} = f_{-2} = 0$):
\begin{equation}
\begin{array}{lcl}
\vspace{1.5mm}
  f_{1} & = & 0, \\

\vspace{1.5mm}
  f_{2} & = & - \displaystyle\frac{f_{0}}{3}\, \bigl[ a_{0} + 2m f_{0} + f_{0}^{2} \bigr], \\

\vspace{1.5mm}
  f_{3} & = & -\, \displaystyle\frac{a_{1} f_{0}}{3}, \\

\vspace{0.5mm}
  f_{n+2} & = &
  -\, \displaystyle\frac{f_{2} f_{n}}{(n+2)\, f_{0}}\: +\:
  \displaystyle\frac{1}{(n+2)\, f_{0}}\,
  \displaystyle\sum\limits_{s=0}^{n}
  \Bigl[
    \displaystyle\frac{(s+1)\, (3n - 5s - 1)}{4}\, f_{s+1}\, f_{n-s+1}\; - \\
  & - &
    f_{s}\, \bigl( a_{0} f_{n-s} + a_{1} f_{n-s-1} + a_{2} f_{n-s-2} \bigr) -
    \displaystyle\sum\limits_{p=0}^{s}
    \displaystyle\sum\limits_{t=0}^{p}
      f_{t}\, f_{p-t}\, f_{s-p}\, f_{n-s} -
    2m \displaystyle\sum\limits_{t=0}^{s} f_{t} f_{s-t}\, f_{n-s}
  \Bigr].
\end{array}
\label{eq.3.2.3}
\end{equation}
Note that there is arbitrary choice in definition for $f_{0}$
(our analysis has shown that variations of this coefficient change strongly the potential $V_{1}$ for larger $r$).
So, for it definition, it needs to introduce a new additional condition (see further in text).

\subsection{Representation of solutions via dimensionless variable
\label{sec.3.3}}

In dimensionless variable~(\ref{eq.2.3.1}), for the spherically symmetric nuclei in the ground state
we have the following change of potential
\begin{equation}
   V_{1}(\xi_{r}) = V(\xi_{r}) - \varepsilon - m
\label{eq.3.3.2}
\end{equation}
and Eq.~(\ref{eq.3.1.6}) is transformed into
\begin{equation}
\begin{array}{cc}
  4\, \bar{V}_{\xi}(\xi_{r})  V_{1}^{2} =
  - 2\, V_{1} V_{1}^{\prime\prime} +
  3\,(V_{1}^{\prime})^{2} -
  \displaystyle\frac{4}{\xi_{r}}\, V_{1} V_{1}^{\prime}\, -
  4\, a_{r}^{2} V_{1}^{4} - 8\, a_{r}^{2} V_{1}^{3}\, m.
\end{array}
\label{eq.3.3.6}
\end{equation}
Solution of this equation in the dimensionless variable $\xi_{r}$ obtains form:
\begin{equation}
\begin{array}{lcl}
\vspace{1mm}
  V_{1}(\xi_{r}) & = & \displaystyle\sum\limits_{n=0} f_{\xi,n} \xi_{r}^{n}, \\
\vspace{1mm}
  V_{1}^{\prime}(\xi_{r}) & = & \displaystyle\sum\limits_{n=0} (n+1) f_{\xi,n+1} \xi_{r}^{n}, \\
  V_{1}^{\prime\prime}(\xi_{r}) & = & \displaystyle\sum\limits_{n=0} (n+1)(n+2) f_{\xi, n+2} \xi_{r}^{n},
\end{array}
\label{eq.3.4.1}
\end{equation}
where
($n \ge 2$, we use $f_{\xi,-1} = f_{\xi,-2} = 0$)
\begin{equation}
\begin{array}{lcl}
\vspace{1.5mm}
  f_{\xi, 1} & = & 0, \\

\vspace{1.5mm}
  f_{\xi, 2} & = & - \displaystyle\frac{f_{\xi,0}}{3}\, \bigl[ a_{\xi,0} + 2a_{r}^{2} m f_{\xi,0} + a_{r}^{2} f_{\xi,0}^{2} \bigr], \\

\vspace{1.5mm}
  f_{\xi, 3} & = & -\, \displaystyle\frac{a_{\xi, 1} f_{\xi, 0}}{3}, \\

\vspace{0.5mm}
  f_{n+2} & = &
  -\, \displaystyle\frac{f_{2} f_{n}}{(n+2)\, f_{0}}\: +\:
  \displaystyle\frac{1}{(n+2)\, f_{0}}\,
  \displaystyle\sum\limits_{s=0}^{n}
  \Bigl[
    \displaystyle\frac{(s+1)\, (3n - 5s - 1)}{4}\, f_{s+1}\, f_{n-s+1}\; - \\
  & - &
    a_{r}^{2} \displaystyle\sum\limits_{p=0}^{s}
    \displaystyle\sum\limits_{t=0}^{p}
      f_{t}\, f_{p-t}\, f_{s-p}\, f_{n-s} -
    2a_{r}^{2} m \displaystyle\sum\limits_{t=0}^{s} f_{t} f_{s-t}\, f_{n-s} -
    f_{s}\, \bigl( a_{\xi,0} f_{n-s} + a_{\xi,1} f_{n-s-1} + a_{\xi,2} f_{n-s-2} \bigr)
  \Bigr]
\end{array}
\label{eq.3.4.3}
\end{equation}
and we use the potential $\bar{V}_{\xi}$ in a form
\begin{equation}
  \bar{V}_{\xi} (\xi) = a_{\xi,0} + a_{\xi,1} \xi_{r} + a_{\xi,2} \xi_{r}^{2},
\label{eq.3.4.2}
\end{equation}
where $a_{\xi,0}$, $a_{\xi,1}$ and $a_{\xi,2}$ are parameters, defined by the microscopic shell model of nucleus,
$f_{\xi,0}$ is arbitrary.

\subsection{Approach of Numerov's type for solution
\label{sec.3.5}}

We develop discrete solution approach to find the unknown potential.
At point $\xi_{r,n}$ we have $v_{1,n} = V_{1}(\xi_{r,n})$ and write derivatives:
\begin{equation}
\begin{array}{cc}
  V_{1}^{\prime} (\xi_{r,n}) = v_{1,n}^{\prime} = \displaystyle\frac{v_{1,n} - v_{1,n-1}}{\Delta}, &
  V_{1}^{\prime\prime} (\xi_{r,n}) = v_{1,n}^{\prime\prime} = \displaystyle\frac{v_{1,n+1} - 2v_{1,n} + v_{1,n-1}}{\Delta^{2}}.
\end{array}
\label{eq.3.5.1}
\end{equation}
We rewrite Eq.~(\ref{eq.3.3.6}) as
\begin{equation}
\begin{array}{cc}
  4\, \bar{v}_{\xi,n}  v_{1,n}^{2} =
  - 2\, v_{1,n} \displaystyle\frac{v_{1,n+1} - 2v_{1,n} + v_{1,n-1}}{\Delta^{2}} +
  3\, \Bigl( \displaystyle\frac{v_{1,n} - v_{1,n-1}}{\Delta} \Bigr)^{2} -
  \displaystyle\frac{4}{\xi_{r}}\, v_{1,n} \Bigl( \displaystyle\frac{v_{1,n} - v_{1,n-1}}{\Delta} \Bigr)\, -
  4\, a_{r}^{2} v_{1,n}^{4} - 8\, a_{r}^{2} v_{1,n}^{3}\, m.
\end{array}
\label{eq.3.5.2}
\end{equation}
Solving this equation, we obtain:
\begin{equation}
\begin{array}{ccl}
  v_{1,n+1}
  & = &
  2\, \Delta^{2}\, \bar{v}_{\xi,n}  v_{1,n} +
  \displaystyle\frac{7}{2}\,v_{1,n} -
  4v_{1,n-1} +
  \displaystyle\frac{3}{2v_{1,n}}\,v_{1,n-1}^{2} -
  \displaystyle\frac{2\Delta\, v_{1,n}}{\xi_{r,n}} +
  \displaystyle\frac{2\Delta\, v_{1,n-1}}{\xi_{r,n}} -
  2\, \Delta^{2} a_{r}^{2} v_{1,n}^{3} -
  4\, \Delta^{2} a_{r}^{2} v_{1,n}^{2}\, m.
\end{array}
\label{eq.3.5.3}
\end{equation}
We choose starting conditions (according to presentation (\ref{eq.3.4.1})):
\begin{equation}
\begin{array}{ccl}
  v_{1,0} = f_{\xi,0}, &
  v_{1,0}^{\prime} = f_{\xi,1} = 0.
\end{array}
\label{eq.3.5.4}
\end{equation}

\section{Comparison of the potential of the oscillator model with the found potential $V_{1}$
\label{sec.4}}

\subsection{Spherically symmetric linear harmonic oscillator
\label{sec.4.1}}

The radial Schr\"{o}dinger equation for linear harmonic oscillator with a frequency $w_{r}$
corresponding to the spherically symmetric nuclei in the ground state
has a form
($n \ge l+1$, $l=0$)%
\footnote{The parameter $a_{r}$ is chosen in such a form in order to correspond completely to formalism of papers [Steshenko]
(for example, this choice is different from approach given in review [Tang] on Resonating group approach.
This allows to use and apply results and estimations of different parameters and characteristics for nuclei given in [Steshenko, Filipov, Vasilevsky]
in our further analysis).}
\begin{equation}
\begin{array}{lcl}
  \Bigl\{
    - \displaystyle\frac{\hbar^{2}}{2m}\, \displaystyle\frac{d^{2}}{dr^{2}} +
    v_{\rm shell} (r)
  \Bigr\}\, \chi_{n}(r) =
  E_{n} \chi_{n}(r), &
  v_{\rm shell} (r) = \displaystyle\frac{m w_{r}^{2} r^{2}}{2}, &
  a_{r}^{2} = \displaystyle\frac{\hbar}{mw_{r}},
\end{array}
\label{eq.4.2.6}
\end{equation}
where solution is (see Ref.~\cite{Landau.v3.1989}, Eqs.~(23.6)--(23.13) in pages~97--99)
\begin{equation}
\begin{array}{lcll}
  \chi_{n}(r) & = &
    \displaystyle\frac{1}{\sqrt{2^{n} n!}}\,
    \Bigl( \displaystyle\frac{1}{\pi\,a_{r}^{2}} \Bigr)^{1/4}
    \exp\Bigl( -\displaystyle\frac{r^{2}}{2\,a_{r}^{2}} \Bigr)\,
    H_{n} \Bigl( \displaystyle\frac{r}{a_{r}} \Bigr), &
  E_{n} = \hbar\, w_{r}\, \Bigl(n + \displaystyle\frac{1}{2} \Bigr)
\end{array}
\label{eq.4.2.7}
\end{equation}
and $H_{n}\, (x)$ are Hermitian Polynomials.
One can rewrite Eq.~(\ref{eq.4.2.6}) as
\begin{equation}
\begin{array}{lcll}
  \Bigl\{ - \displaystyle\frac{d^{2}}{dr^{2}} + V_{n} (r) \Bigr\}\, \chi_{n} (r) = 0, &
  V_{n} (r) =
    \displaystyle\frac{2m}{\hbar^{2}}\,
    \Bigl\{ \displaystyle\frac{m w_{r}^{2} r^{2}}{2} - E_{n} \Bigr\}.
\end{array}
\label{eq.4.2.8}
\end{equation}
This equation should be equivalent to Eq.~(\ref{eq.2.2.4}).
So, we impose conditions of $\bar{V}(r) = V_{n}(r)$, $h(r) = \chi(r)$ (both equations have the same variable $r$):
\begin{equation}
\begin{array}{lcl}
\vspace{1mm}
  h_{n} (r) = \chi_{n} (r), &
  \bar{V}(r) =
  V_{n} (r) =
    \displaystyle\frac{2m}{\hbar^{2}}\,
    \Bigl\{ \displaystyle\frac{m w_{r}^{2} r^{2}}{2} - E_{n} \Bigr\}.
\end{array}
\label{eq.4.2.9}
\end{equation}
In particular, for the ground state we have
[$n = 1$, $H_{1}(x) = 2x$,
see Ref.~\cite{Landau.v3.1989}, Eq.~(a.6) in page~781]:
\begin{equation}
\begin{array}{lcl}
  h_{1} (r) =
  \displaystyle\frac{\sqrt{2}}{\pi^{1/4}\, \sqrt{a_{r}}}\,
    \exp\Bigl(- \displaystyle\frac{r^{2}}{2\, a_{r}^{2}} \Bigr)\,
    \displaystyle\frac{r}{a_{r}}, &
  h_{1}^{\prime} (r) = \Bigl( - \displaystyle\frac{r}{a_{r}^{2}} + \displaystyle\frac{1}{r} \Bigr)\, h_{1} (r).
\end{array}
\label{eq.4.2.10}
\end{equation}

\subsection{Calculations for the ground state for $^{4}{\rm He}$
\label{sec.4.3}}

\begin{figure}[htbp]
\centerline{%
\includegraphics[width=86mm]{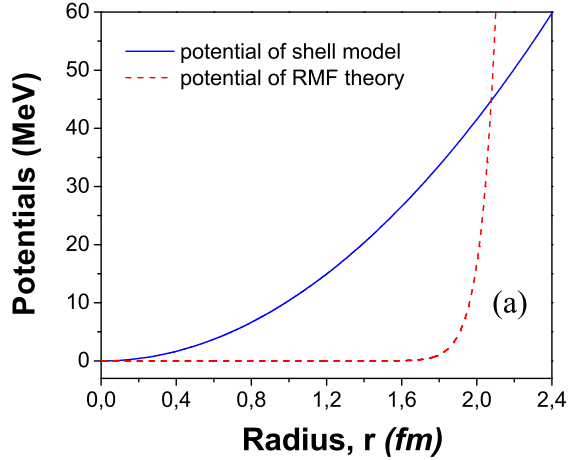}
}
\vspace{-3mm}
\caption{\small (Color online)
The potential $v_{\rm shell} (r)$ of the shell model defined in Eq.~(\ref{eq.4.2.6}) (see solid blue line),
and the potential of RMF theory $V_{1}(r)$ defined by Eq.~(\ref{eq.3.4.1}) and expressed via variable $r$ (see dashed red line),
calculated for the ground state of $^{4}{\rm He}$.
One can see that both potentials provide only bound states. However, their shapes are essentially different.
\label{fig.4.1}}
\end{figure}

In frameworks of the oscillator model, for the ground state of $^{4}{\rm He}$ we have
\begin{equation}
\begin{array}{lcl}
  a_{r} = 1.2\; {\rm fm}, &
  w_{r} = \displaystyle\frac{\hbar}{m a_{r}^{2}} = 29.892\; {\rm MeV}, &
  E_{1} =\displaystyle\frac{3\, \hbar w_{r}}{2} = 44.8390657954729\; {\rm MeV}.
\end{array}
\label{eq.4.3.1}
\end{equation}
For determination of the unknown $f_{0}$ we introduce such a new condition:
\emph{The ground states should be equivalent in frameworks of oscillator model and RMF theory}.
In practical calculations,
we look for value for $f_{0}$, when the potential $V_{1}(r_{0})$ given by Eqs.~(\ref{eq.3.4.1})
coincides with the potential of the shell model $v_{\rm shell}(r_{0})$
%
%
which corresponds to energy $E_{1}$ of the ground state (this is radial point of intersection, $r_{0}=\sqrt{3}\,a_{r} = 2.078460$~fm).
On such a basis, we find (in calculations we use $N=26$ as upper limit in summation~(\ref{eq.3.4.1})):%
\footnote{We obtain the same results for $f_{0}$ and $V_{1}(r_{0})$,
if to use series presentation~(\ref{eq.3.2.1}) with coefficients~(\ref{eq.3.2.3}) for the potential $V_{1}(r)$
and presentation~(\ref{eq.3.2.2}) for the potential $\bar{V}(r)$.}
\begin{equation}
\begin{array}{lcl}
  f_{0} = 3.63624108\; 10^{-8}\; {\rm MeV}, &
  V_{1} (r_{0}) = 44.8390655772297\; {\rm MeV}, &
  \bar{V}(r_{0}) = 0\; {\rm MeV}^{2}.
\end{array}
\label{eq.4.3.2}
\end{equation}
In Fig.~\ref{fig.4.1} we present the potentials $v_{\rm shell} (r)$ and $V_{1}(r)$ calculated by the approach above.
From such a picture we see an essential difference between these two potentials. This is the first confirmation of not equivalence in quantum descriptions of nuclear properties and processes provided by the shell model and RMF theory.
On such a basis, one can see that difference in quantum description of nuclear processes given by these two potentials given by the shell model and
RMF theory should be essential in some cases.
Also we obtain:
\begin{equation}
\begin{array}{lll}
  a_{\xi,0} = - \displaystyle\frac{2 E_{1}}{\hbar w_{r}} =
    - \displaystyle\frac{2}{\hbar w_{r}}\, \displaystyle\frac{3}{2}\, \hbar w_{r} = - 3, &
  a_{\xi,1} = 0, &
  a_{\xi,2} = 1.
\end{array}
\label{eq.4.3.3}
\end{equation}

\section{Wave functions of nucleons and nucleon densities
\label{sec.5}}

\subsection{Wave function and density calculations
\label{sec.5.1}}

We shall write a final solution for the wave function of nucleons.
Substituting solution (\ref{eq.3.2.1}) for $V_{1}(r)$ to Eqs.~(\ref{eq.2.2.1}), (\ref{eq.2.2.6}), we obtain
\begin{equation}
\begin{array}{lcl}
  g\,(r) & = & n_{0}\, \sqrt{\Bigl|f_{0} + \displaystyle\sum\limits_{n=1} f_{n} r^{n} \Bigr|} \cdot \displaystyle\frac{h\,(r)}{r}, \\
  f\,(r) & = &
    -\, {\rm sign}\, \bigl[ V_{1}(r) \bigr]\;
    \displaystyle\frac{n_{0}}{r\: \sqrt{|f_{0} + \displaystyle\sum\limits_{n=1}^{N} f_{n} r^{n}|}}\;
    \biggl\{
      \Bigl[
        \displaystyle\frac{f_{1} + \displaystyle\sum\limits_{n=1} (n+1) f_{n+1} r^{n}}
          {2\, (f_{0} + \displaystyle\sum\limits_{n=1} f_{n} r^{n})} -
        \displaystyle\frac{1}{r}\;
      \Bigr]\; h\,(r) +
      h^{\prime}\,(r)
    \biggr\}.
\end{array}
\label{eq.5.1.6}
\end{equation}
For the ground state of $^{4}{\rm He}$ we have
\begin{equation}
\begin{array}{lcl}
  g\,(r) = n_{0}\, \sqrt{\bigl|V_{1}(r)\bigr|} \cdot \displaystyle\frac{h_{1}\,(r)}{r}, &
  f\,(r) =
    {\rm sign}\, \bigl[ V_{1}(r) \bigr]\;
    \displaystyle\frac{n_{0}}{\sqrt{\bigl| V_{1}(r) \bigr|}}\;
    \biggl\{
      - \displaystyle\frac{V_{1}^{\prime}(r)}{2\, V_{1}(r)} +
      \displaystyle\frac{r}{a_{r}^{2}}
    \biggr\} \cdot
    \displaystyle\frac{h_{1}\,(r)}{r}.
\end{array}
\label{eq.5.1.9}
\end{equation}
%
We find the density $\rho_{v} (\mathbf{r})$ defined in Eq.~(42) in Ref.~\cite{Ring.1996.PPNP}.
For the $^{4}{\rm He}$ nucleus in the ground state we obtain
($n = 1$, $l = 0$, $j = \displaystyle\frac{1}{2}$, $m = 0$, $M = 1/2$)
\begin{equation}
\begin{array}{lllll}
  \rho_{v} (\mathbf{r}) & = &
%
%
  n_{0}^{2}\,
  \biggl\{
    \Bigl|f_{0} + \displaystyle\sum\limits_{n=1} f_{n} r^{n} \Bigr| +
    \displaystyle\frac{1}{\bigl| f_{0} + \displaystyle\sum\limits_{n=1} f_{n} r^{n} \bigr|}\;
    \biggl|
      - \displaystyle\frac{f_{1} + \displaystyle\sum\limits_{n=1} (n+1) f_{n+1} r^{n}}
        {2\, (f_{0} + \displaystyle\sum\limits_{n=1} f_{n} r^{n})} +
      \displaystyle\frac{r}{a_{r}^{2}}
    \biggr|^{2}
  \biggr\}\,
  \displaystyle\frac{h_{1}^{2}(r)}{r^{2}}.
\end{array}
\label{eq.5.2.7}
\end{equation}
This solution can be represented in form of series as
\begin{equation}
\begin{array}{lllll}
  \rho_{v} (\mathbf{r}) & = &
  n_{0}^{2}\,
  \displaystyle\frac{h^{2}\,(r)}{r^{2}}\,
  \displaystyle\sum\limits_{n=0} c_{n} r^{n}.
\end{array}
\label{eq.5.2.8}
\end{equation}

\subsection{Calculations of coefficients $c_{n}$
\label{sec.5.3}}

We find the unknown coefficients $c_{n}$ in Eq.~(\ref{eq.5.2.8}).
We combine Eqs.~(\ref{eq.5.2.7}) and (\ref{eq.5.2.8}) and obtain the following equation
\begin{equation}
\begin{array}{lllll}
  \displaystyle\sum\limits_{n=0} c_{n} r^{n} =
    \Bigl|f_{0} + \displaystyle\sum\limits_{n=1} f_{n} r^{n} \Bigr| +
    \displaystyle\frac{1}{\Bigl| f_{0} + \displaystyle\sum\limits_{n=1} f_{n} r^{n} \Bigr|}\;
    \biggl|
      - \displaystyle\frac{f_{1} + \displaystyle\sum\limits_{n=1} (n+1) f_{n+1} r^{n}}
        {2\, (f_{0} + \displaystyle\sum\limits_{n=1} f_{n} r^{n})} +
      \displaystyle\frac{r}{a_{r}^{2}}
    \biggr|^{2}.
\end{array}
\label{eq.5.3.1}
\end{equation}
Solving this equation, we obtain the following values for the coefficients
(at $n \ge 2$, see App.~\ref{sec.app.4}, for details):
\begin{equation}
\begin{array}{lllll}
  c_{0} = f_{0} + \displaystyle\frac{f_{1}^{2}}{4f_{0}^{3}}, \\
  c_{1} = A_{1} f_{0}^{-3} - 3 f_{0}^{-1} f_{1}\, c_{0}, \\
  c_{n} =
  A_{2} f_{0}^{-3} -
  f_{0}^{-3} \displaystyle\sum\limits_{s=1}^{n}
    \displaystyle\sum\limits_{p=0}^{s}
    \displaystyle\sum\limits_{t=0}^{p} f_{t}\, f_{p-t}\, f_{s-p}\, c_{n-s},
\end{array}
\label{eq.5.3.2}
\end{equation}
where
\begin{equation}
\begin{array}{lllll}
  A_{1} & = &
  4\, f_{0}^{3} f_{1} +
  f_{1} f_{2} -
  \displaystyle\frac{1}{a_{r}^{2}} f_{0} f_{1}, \\

  A_{2} & = &
  \displaystyle\sum\limits_{s=0}^{n}
    \displaystyle\sum\limits_{p=0}^{s}
    \displaystyle\sum\limits_{t=0}^{p}
      f_{t}\, f_{p-t}\, f_{s-p}\, f_{n-s} +
  \displaystyle\frac{1}{4}\,
    \displaystyle\sum\limits_{s=0}^{n}
    (s+1) (n-s+1)\, f_{s+1} f_{n-s+1} - \\
  & - &
  \displaystyle\frac{1}{a_{r}^{2}}
    \displaystyle\sum\limits_{s=0}^{n-1} (s+1)\, f_{s+1} f_{n-s-1} +
  \displaystyle\frac{1}{a_{r}^{4}}\, \displaystyle\sum\limits_{s=0}^{n-2} f_{s} f_{n-s-2}.
\end{array}
\label{eq.5.3.3}
\end{equation}
From the obtained solutions we see, that these coefficients do not depend explicitly on factor $n_{0}$.
However, such a dependence is included in calculations of coefficients $f_{n}$ in Eqs.~(\ref{eq.3.2.3}).

\begin{figure}[htbp]
\centerline{\includegraphics[width=86mm]{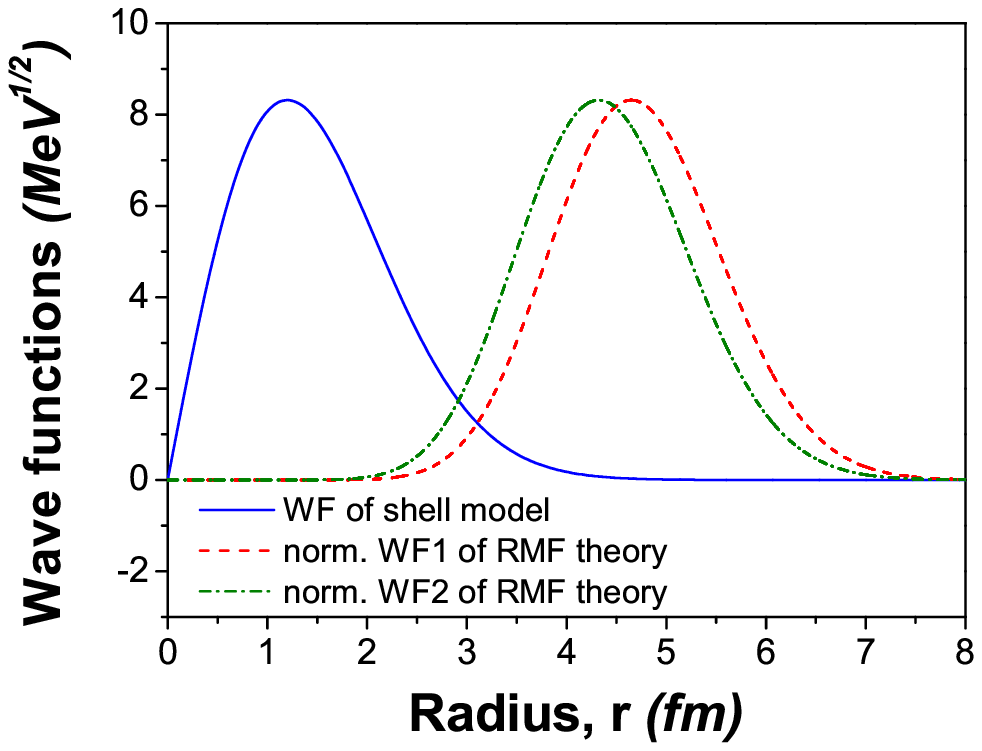}
\hspace{-5mm}\includegraphics[width=86mm]{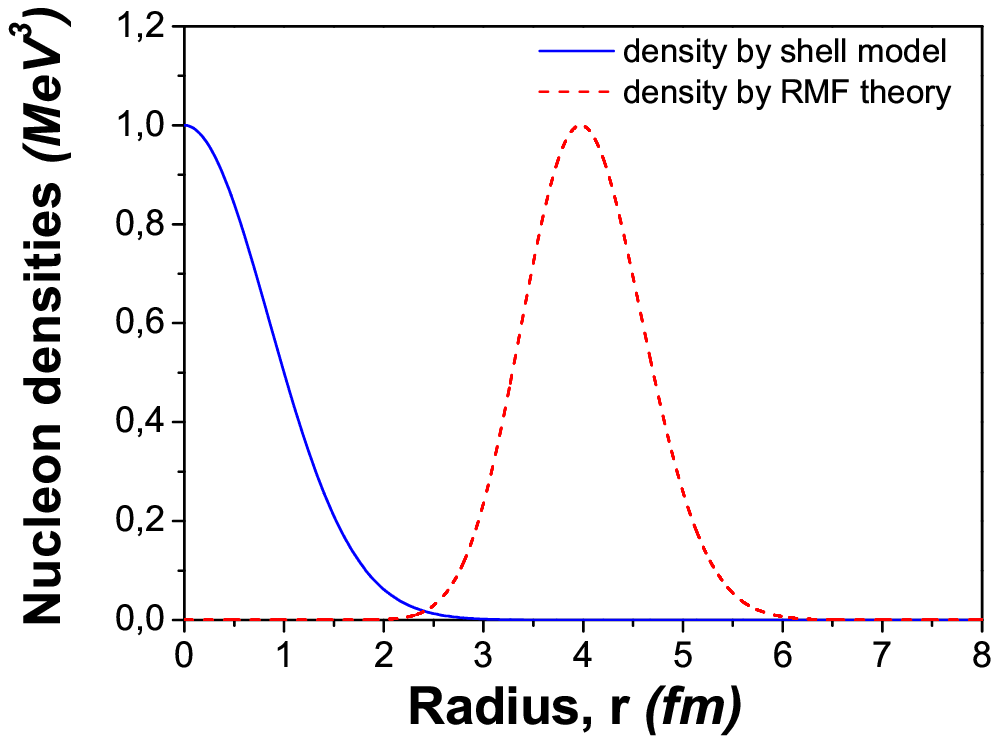}}
\vspace{-3mm}
\caption{\small (Color online)
The nucleon radial wave functions and the densities of the shell model and RMF theory
for the $^{4}{\rm He}$ nucleus in the ground state.
Panel a:
The wave function $h_{1}(r)$ of the shell model defined in Eq.~(\ref{eq.4.2.10}) at $n=1$ for the ground state (see solid blue line),
the wave function $r\, g(r)$ of the RMF theory defined in Eq.~(\ref{eq.5.1.6}) and re-scaled on maximum of the function $h_{1}(r)$ (see dashed red line), and
the wave function $r\, f(r)$ of the RMF theory defined in Eq.~(\ref{eq.5.1.6}) and re-scaled on maximum of the function $h_{1}(r)$ (see dash-dotted green line).
One case see difference in the shapes of theses wave functions:
we observe a clear radial distance between maximums of the function of the shell model [$h_{1}(r_{\rm max})=8.31769\,{\rm MeV}^{1/2}$ at $r_{\rm max}=1.2024$~fm]
and functions of RMF theory
[$r_{\rm max}\, g(r_{\rm max})=162.29505\,{\rm MeV}^{1/2}$ at $r_{\rm max}=4.64930$~fm, and
$r_{\rm max}\, f(r_{\rm max})=8.73757 \cdot 10^{8}\,{\rm MeV}^{1/2}$ at $r_{\rm max}=4.32866$~fm].
From here we conclude that a probability of nucleons described by the wave function $f(r)$ is essentially larger than the probability given by the function~$g(r)$.
So, for more quick calculations we can recommend to use only the function $f(r)$ in analysis of quantum properties of this nucleus.
Panel b:
The nucleon density of the shell model defined as $\rho_{v} (r) = h_{1}^{2}\,(r) / r^{2}$
(see solid blue line),
the nucleon density $\rho_{v}(r)$ of the RMF theory defined in Eq.~(\ref{eq.5.2.7}) (see dashed red line, its maximum is at $3.96794$~fm)
(the densities are re-scaled on unite at maximums).
One can see clear difference between maximums of both lines which describes different nucleon distributions inside the nucleus given by the shell model and RMF theory.
Such a difference is explained by appearance of the additional factor $|V_{1}(r)|^{1/2}$ in the radial wave function $g(r)$ and
the corresponding factor $|V_{1}|^{-1/2}\, ( - V_{1}^{\prime}/(2V_{1}) + r a_{r}^{-2} )$ in the radial wave function $f(r)$ [see Eqs.~(\ref{eq.5.1.9})].
\label{fig.5.1}}
\end{figure}

In Fig.~\ref{fig.5.1} we present our calculations of the wave functions and densities provided by the shell model and RMF theory
for the ground state of $^{4}{\rm He}$.
One can see that these two approaches give essentially different probabilities of nucleons in dependence on the radial distance, which provide principally different physical description of quantum picture of nucleon distribution inside this nucleus.
From such a picture,
the nucleon densities given by RMF theory contradict to knowledge about distribution of the proton and neutron densities inside the nuclei obtained from experimental study of
scattering of high energetic electrons off nuclei,
spectroscopic information about transitions in muon atoms,
isotopic and isomer shifts in the optical spectra
(see Refs.~\cite{De_Jager.1974.ADNDT,De_Vries.1987.ADNDT,Angeli.2004.ADNDT}).
This strongly indicate that the functions $g$ and $f$ have no sense as the wave functions provided by quantum physics.
At the same time, the function $h$ which is connected with these two functions above, provides us a proper description of quantum properties of nucleons inside the $^{4}{\rm He}$ nucleus.

\begin{figure}[htbp]
\centerline{\includegraphics[width=86mm]{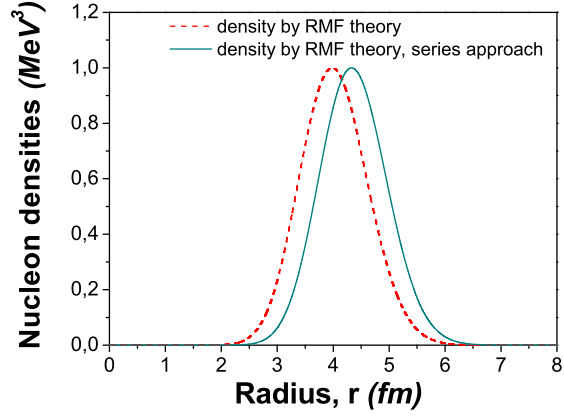}}
\vspace{-3mm}
\caption{\small (Color online)
The nucleon densities of the RMF theory.
The density $\rho_{v}(r)$ is defined in Eq.~(\ref{eq.5.2.7}) (see dashed red line),
the density $\rho_{v}(r)$ is defined in series expansion approach~(\ref{eq.5.2.8}) with coefficients determined by Eqs.~(\ref{eq.5.3.2}) and ~(\ref{eq.5.3.3}) (see solid dark cyan line)
(the densities are re-normalized on unite at maximums).
A difference between shapes is explained by the accuracy of the series expansion approach
(that characterizes accuracy of determination of coefficients in Eqs.~(\ref{eq.5.3.2}) and ~(\ref{eq.5.3.3})).
\label{fig.5.2}}
\end{figure}

\section{Solution of equations for mesons
\label{sec.6}}

\begin{figure}[htbp]
\centerline{\includegraphics[width=86mm]{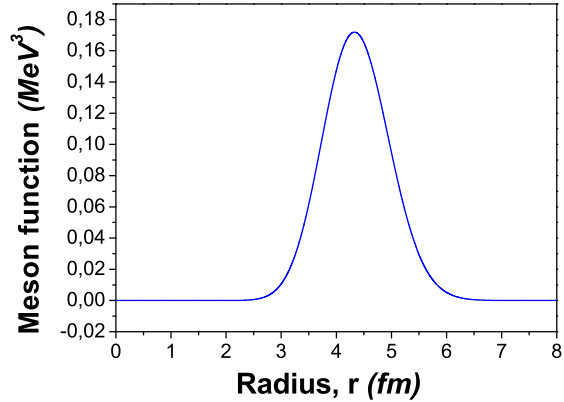}}
\vspace{-3mm}
\caption{\small (Color online)
The meson radial wave function $R_{w}(r)$ defined by Eqs.~(\ref{eq.6.2}) and (\ref{eq.6.5})
where the coefficients $c_{w,n}$ are determined by Eqs.~(\ref{eq.6.8})
[maximum of this function is $R_{w}(r_{\rm max}) = 0.17198\, {\rm MeV}^{3}$ at $r_{\rm max} = 4.32866$~fm,
we use $c_{w,0}=1$ in calculations].
One can observe similarity (after re-scaling) of its shape with the shape of the nucleon density given in Figs.~\ref{fig.5.1}(b) and \ref{fig.5.2}.
\label{fig.6.1}}
\end{figure}

In this section we shall find solution of the system of equations for mesons $w$ defined in Eq.~(38) in Ref.~\cite{Ring.1996.PPNP}.
\begin{equation}
  (- \triangle + m_{w})\, w^{0} (\mathbf{r}) = g_{w}\, \rho_{v} (\mathbf{r}).
\label{eq.6.1}
\end{equation}
We shall rewrite the nucleon density in the right part of this equation in form of series expansion as (\ref{eq.5.2.8})
and we use the following form for the unknown function
\begin{equation}
\begin{array}{lcl}
  w^{0}(r, \theta, \phi) = R_{w}(r)\, Y_{lm}(\theta, \phi), &
  R_{w}(r) = \displaystyle\frac{\chi_{w}(r)}{r}.
\end{array}
\label{eq.6.2}
\end{equation}
Then the left part of Eq.~(\ref{eq.6.1}) is transformed into
\begin{equation}
\begin{array}{lcl}
  \Bigl\{
    - \displaystyle\frac{d^{2}}{dr^{2}} +
    \displaystyle\frac{l(l+1)}{r^{2}} +
    m_{w}^{2}
  \Bigr\}\, \chi_{w}(r)\, Y_{lm}(\theta, \phi) =
  g_{w}\, \rho_{v} (\mathbf{r})\; r.
\end{array}
\label{eq.6.3}
\end{equation}
As the density function $\rho_{v} (\mathbf{r})$ has no angular part, this equation has solution only in case of $l=0$ (we are interesting in such a case).
So, Eq.~(\ref{eq.6.3}) is transformed to one-dimensional
\begin{equation}
\begin{array}{lcl}
  \Bigl\{ - \displaystyle\frac{d^{2}}{dr^{2}} + m_{w}^{2} \Bigr\}\, \chi_{w}(r) =
  g_{w}\, \rho_{v} (\mathbf{r})\; r.
\end{array}
\label{eq.6.4}
\end{equation}
We look for the unknown function $\chi_{w}(r)$ in form of the series expansion as
\begin{equation}
\begin{array}{lllll}
 \chi_{w} (r) & = &
  n_{0}^{2}\,
  \displaystyle\frac{h^{2}(r)}{r}\,
  \displaystyle\sum\limits_{n=0} c_{w,n} r^{n},
\end{array}
\label{eq.6.5}
\end{equation}
where $c_{w,n}$ are new unknown coefficients.
We calculate derivatives, rewrite Eq.~(\ref{eq.6.4}) as
%
\begin{equation}
\begin{array}{lllll}
\vspace{0.5mm}
  - \chi_{w}^{\prime\prime} (r) + m_{w}^{2} \chi_{w} & = &
  - n_{0}^{2}\,
  \displaystyle\frac{h^{2}(r)}{r}\;
  \biggl\{
    \displaystyle\frac{2\, c_{w,1}}{r} +
    6\, \bigl[c_{w,2} - a_{r}^{-2}c_{w,0} \bigr] +
    \bigl[12\, c_{w,3} - 10a_{r}^{-2}c_{w,1} \bigr]\, r\; - \\
  & + &
    \displaystyle\sum\limits_{n=2}
      \Bigl[
         4a_{r}^{-4}c_{w,n-2} - 2a_{r}^{-2} (2n+3)\, c_{w,n} + (n+2)\,(n+3)\, c_{w,n+2}
      \Bigr]\, r^{n}
  \biggr\}\; +\;

  m_{w}^{2}\, n_{0}^{2}
    \displaystyle\frac{h^{2}(r)}{r}
    \displaystyle\sum\limits_{n=0} c_{w,n} r^{n}\; = \\

  & = &
  g_{w}\, \rho_{v} (\mathbf{r})\; r =
  g_{w}\; n_{0}^{2}\,
    \displaystyle\frac{h^{2}\,(r)}{r}\,
    \displaystyle\sum\limits_{n=0} c_{n} r^{n}.
\end{array}
\label{eq.6.7}
\end{equation}
and find the unknown coefficients (see App.~\ref{sec.app.5}, for details)
\begin{equation}
\begin{array}{lllll}
\vspace{2mm}
  c_{w,1} = 0, \\

\vspace{2mm}
  c_{w,2} = \Bigl( a_{r}^{-2} + \displaystyle\frac{m_{w}^{2}}{6} \Bigr)\,c_{w,0} - \displaystyle\frac{g_{w}}{6}\, c_{0}, \\

\vspace{2mm}
  c_{w,3} = - \displaystyle\frac{g_{w}}{12}\, c_{1}, \\

  c_{w,n+2} =
    - \displaystyle\frac{4a_{r}^{-4}}{(n+2)\,(n+3)}\, c_{w,n-2} +
    \displaystyle\frac{2a_{r}^{-2} (2n+3) + m_{w}^{2}}{(n+2)\,(n+3)}\, c_{w,n} -
    \displaystyle\frac{g_{w}}{(n+2)\,(n+3)}\, c_{n}.
\end{array}
\label{eq.6.8}
\end{equation}
From such solutions we see that there is arbitrary choice in determination of $c_{w,0}$.
In Fig.~\ref{fig.6.1} we present our calculation of the radial component of the meson wave function, $R_{w}(r)$,
given by Eqs.~(\ref{eq.6.2}) and (\ref{eq.6.5}).
But in practical calculations, varying $c_{w,0}$ the shape of the meson wave function $R_{w}(r)$ keeps to be stable (after re-scaling).
At $r \to 0$, the function $\chi_{w}(r) \to 0$ independently on choice of $c_{w,0}$.

\section{Quantum correction to the potential
\label{sec.7}}

The meson function $w_{0}$ was defined in Eqs.~(\ref{eq.6.2}), (\ref{eq.6.5}) and (\ref{eq.6.8}) on the basis of solution~(\ref{eq.5.3.2}) for the wave function $\psi_{\varepsilon jlM}(\mathbf{r})$ in Eqs.~(\ref{eq.2.2.2}) and (\ref{eq.2.2.3})
and the potential $V_{1}$ in Eqs.~(\ref{eq.3.4.1}), (\ref{eq.3.4.3})
which are in direct correspondence with the wave function and potential (\ref{eq.4.2.8}) and (\ref{eq.4.2.9}) of the shell model.
So, the potential $V(r)$ defined in Eq.~(45) in Ref.~\cite{Ring.1996.PPNP}
on the basis of the found meson wave function $w_{0}$, is connected with wave function $\psi_{\varepsilon jlM}(\mathbf{r})$ but this wave functions is not exact solution for this potential in Dirac Eq.~(\ref{eq.2.2.1}).
On the other side, we found the potential $V_{1}$ which is directly connected with the potential of the shell model.
If the meson theory describes quantum properties of nucleus well (like the shell model), then a difference between these two potentials should be as small as possible (tend to zero).
In order to analyze this, we introduce new quantum corrections characterizing difference between these two potentials.
In order to provide the first estimations and realize this idea as clear as possible, we restrict ourselves by the potential $V(r)$
without terms $g_{\rho} \tau_{3} \rho_{3}^{0} (\mathbf{r})$ and $\varepsilon\, A^{0} (\mathbf{r})$ (electromagnetic interactions).
We formulate the corrections, generalizing the potential $V(r)$ defined in Eq.~(45) in Ref.~\cite{Ring.1996.PPNP} as
\begin{equation}
  V_{\rm gen} (r) = g_{\rm quant,w}\, g_{w}\, w^{0}(r) + \Delta V(r) + V_{0},
\label{eq.7.1}
\end{equation}
where we introduced
a new factor $g_{\rm quant,w}$,
a new function $\Delta V(r)$, and
a new constant $V_{0}$.
In order to find these new corrections, we introduce the following requirement:
\emph{Correction $\Delta V(r)$ should be minimal}.
In result, we obtain:
\begin{equation}
\begin{array}{lllll}
\vspace{2mm}
  \Delta V(r) = V_{1}(r) - v_{w}(r), &
  v_{w}(r) = g_{\rm quant, w}\, g_{w}\, w^{0}(r), &
  V_{0} = \varepsilon_{1} + m,
\end{array}
\label{eq.7.2}
\end{equation}
where $\varepsilon_{1}$ is energy of nucleon in the the ground state, and
we introduce a new function $v_{w}(r)$ for more easy demonstration of the meson potential.

\begin{figure}[htbp]
\centerline{\includegraphics[width=86mm]{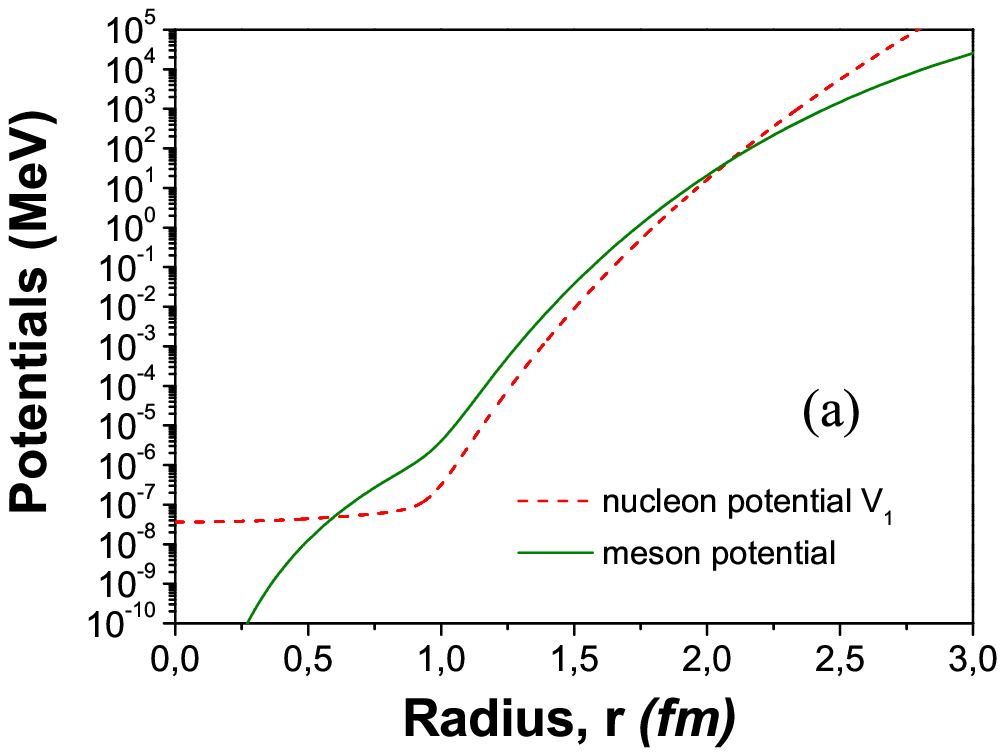}
\hspace{-5mm}\includegraphics[width=86mm]{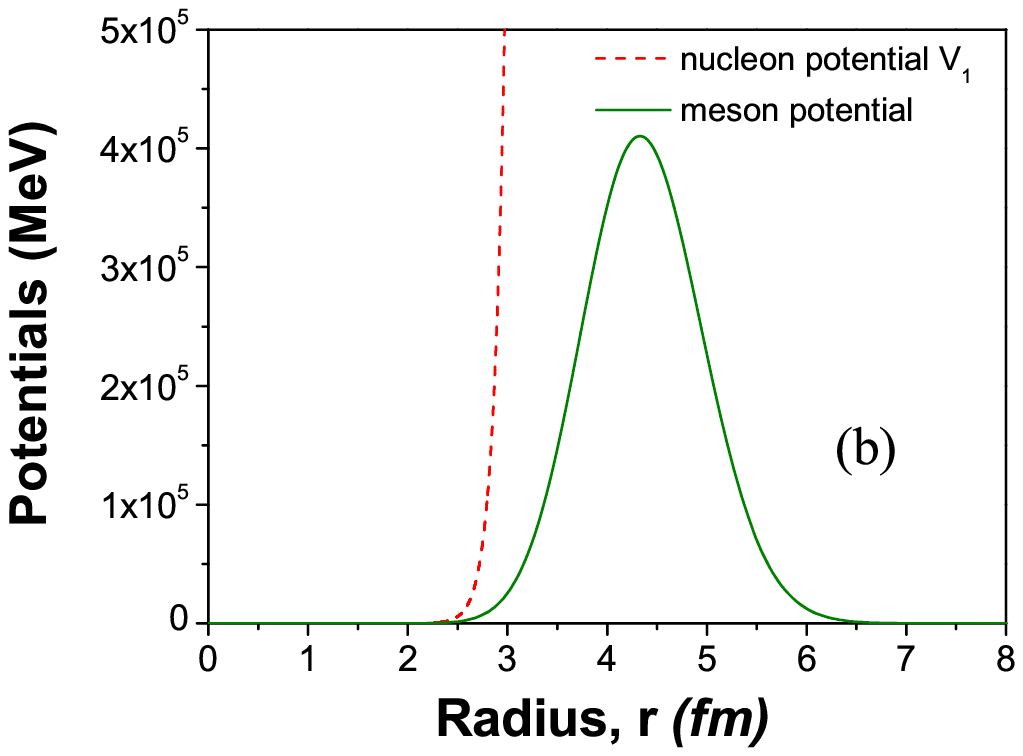}}
\vspace{-3mm}
\caption{\small (Color online)
The nucleon potential $V_{1}(r)$ defined in Eqs.~(\ref{eq.3.4.1}), (\ref{eq.3.4.3}) and the re-normalized meson potential $v_{w} (r)$ defined in Eq.~(\ref{eq.7.2}).
We find an unknown new factor $g_{\rm quant, w}$ from condition of coincidence $V_{1}(r_{0}) = v_{w}(r_{0})$ at the radial point of intersection $r_{0}$ for the ground state
(here, $r_{0}=\sqrt{3}\,a_{r}$, see Fig.~\ref{fig.4.1}).
We obtain $g_{\rm quant, w} = 2.3864127 \cdot 10^{6}$.
This indicates that the initial potential $V_{1}$ should be strongly reinforced in order to minimize maximally the additional quantum correction $\Delta V$. Another conclusion is that without inclusion of this quantum correction, the potential of the RMF theory
defined in Eq.~(45) in Ref.~\cite{Ring.1996.PPNP}
is far from quantum description of properties of the nucleus $^{4}{\rm He}$ given by the shell model.
Panel a:
One can see that difference between these two potentials is very small inside a small radial region at $r \le r_{0}$.
%
Panel b:
For large radial distance, at $r \ge r_{0}$, difference between these two functions is highly increased with increasing of radius $r$. One can see that the quantum correction $\Delta V(r)$
is practically determined only by the potential $V_{1}(r)$ in this region.
A next conclusion is that after inclusion of the new factor $q_{\rm quantum,w}$ (and without adding $\Delta V(r)$), the modified meson potential $v_{w}(r)$ can be successful in description of the quantum properties (on the level of the shell model) for analyzing nucleus $^{4}{\rm He}$ for energies below the ground state or a little above.
It is impossible to describe the quantum properties properly for the energies higher than the ground state energy level without inclusion of the quantum corrections $\Delta V(r)$.
This demonstrates clearly an importance of introduction of the quantum corrections in this paper to the existed RMF theory.
\label{fig.7.1}}
\end{figure}

\begin{figure}[htbp]
\centerline{\includegraphics[width=86mm]{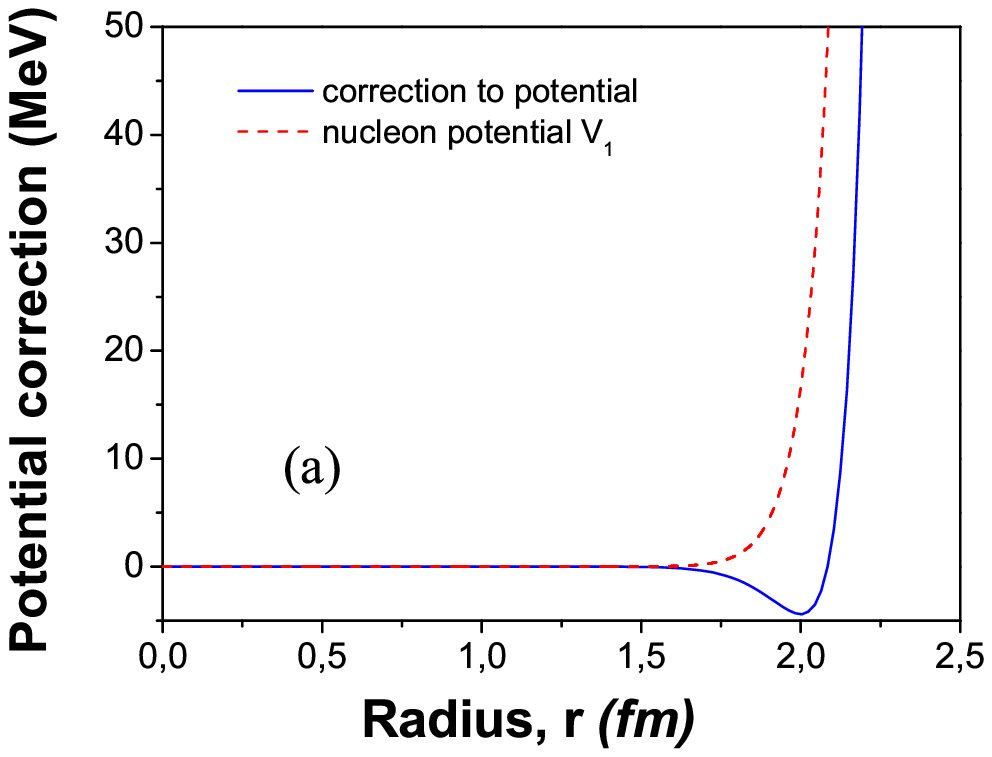}
\hspace{-5mm}\includegraphics[width=86mm]{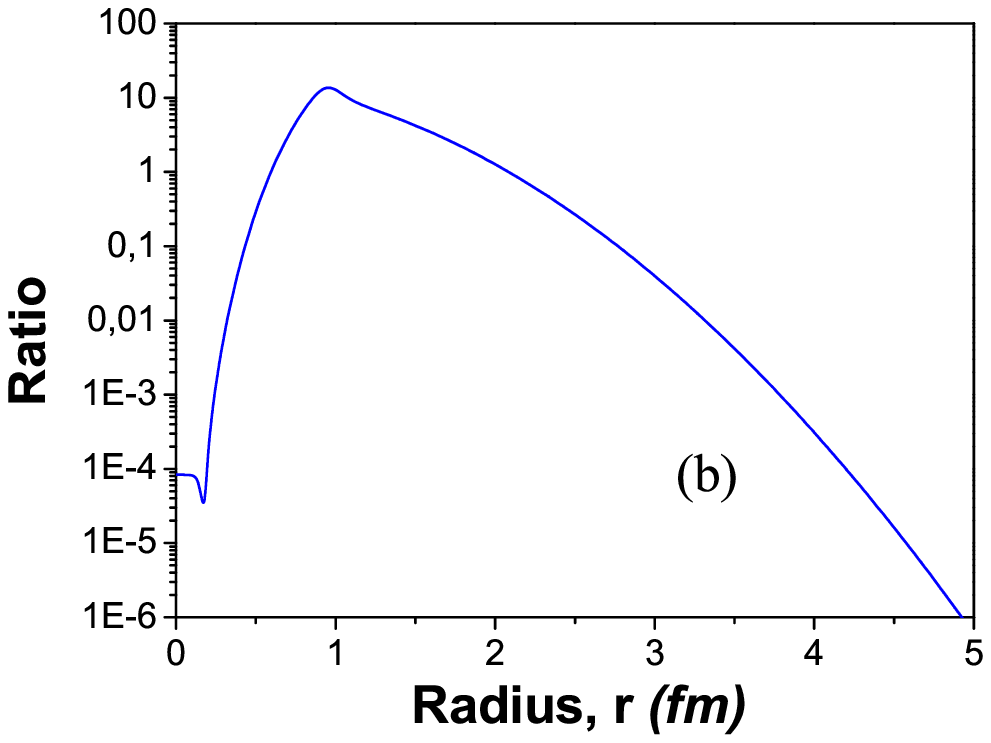}}
\vspace{-3mm}
\caption{\small (Color online)
The quantum correction $\Delta V(r)$ defined in Eq.~(\ref{eq.7.2}) and the potential $V_{1}$.
%
Panel a:
Blue solid line is for the correction $\Delta V$,
red dashed line is for the potential $V_{1}$.
One can see that the quantum correction is very small at $r \le r_{0}$
(this function is negative in this region), but it is increased highly at $r \ge r_{0}$.
%
Panel b:
Ratio $|V_{1}(r) - \Delta V(r)| / V_{1}(r)$.
\label{fig.7.2}}
\end{figure}

Before the calculations of the quantum corrections, at first we analyze
how much closely these two potentials $V_{1}(r)$ and $v_{w}(r)$ can be located, if to vary the factor $g_{\rm quant,w}$.
Results of such calculations are presented in Fig.~\ref{fig.7.1}.
We find that these two functions can be intersected at point $r_{0} = \sqrt{3}\, a_{r}$ (and, so, we use this condition for determination of $g_{\rm quant,w}$).
The final calculated quantum correction $\Delta V(r)$ is presented in Fig.~\ref{fig.7.2} in comparison on the potential $V_{1}$.
One can see that at $r \le r_{0}$ the quantum correction $\Delta V$ is very small, but it is increased highly at increasing $r$ at $r > r_{0}$.

\section{Conclusions
\label{sec.conclusions}}

In this paper, we have compared
the RMF theory~\cite{Ring.1996.PPNP,Ring_Afanasjev.1997.PPNP,Ring.2001.PPNP,Ring_Serra_Rummel.2001.PPNP,
Ring.2007.PPNP,PenaArteaga_Ring.2007.PPNP,Ring.2011.PPNP}
and the microscopic model~\cite{Wildermuth.1977.book,Tang.1978.PR,Tang.1981.lectures,Horiuchi.1977.PTPS}
in description of the quantum properties of the light nuclei in the bound states in the spherically symmetric approximation.
We have obtained an explicit analytical relation between differential equations for the RMF theory and the model of deformed shells, which determine wave functions for nucleons.
On such a basis we perform analysis of correspondence of quantum properties of nuclei described by these wave functions.
Conclusions from analysis on the basis of this approach are the following.
\begin{enumerate}
\item
On the basis of the found relations above, we found the potential $V_{RMF}$ of the RMF theory for nucleons [see Eqs.~(\ref{eq.3.3.2}), (\ref{eq.3.4.1}) and (\ref{eq.3.4.3})] which provides solutions of the wave functions for nucleons, coincident exactly analytically with solutions for the nucleon wave function given by the DOS model.
This fact confirms that it is possible to describe quantum properties of nucleus in framework of the RMF theory exactly the same as in the DOS model.
Note the following.
\begin{enumerate}
\item
However, we find that a difference between the potential $V_{RMF}$ of the RMF theory and the corresponding potential $V_{shell}$ of the DOS model is essential for any nucleus.
On example of the $^{4}{\rm He}$ nucleus in the ground state, we find that
a maximal relative difference between the potentials is for the radius values near zero, 1.6~fm and for region larger 2~fm
(see Fig.~\ref{fig.4.1}).

\item
For determination of the unknown potential $V_{RMF}$ we introduce a series expansion approach [see Eqs.~(\ref{eq.3.4.1}) and (\ref{eq.3.4.3})] which is effective in next calculations of the unknown wave functions for nucleons and densities, solution of equations for mesons. However, this approach provides arbitrary choice of the first unknown coefficient $f_{0}$ (which changes the potential $V_{RMF}$). To resolve this problem, we formulate a new condition:
\emph{The ground states should be equivalent in frameworks of the DOS model and RMF theory}
(see Eq.~(\ref{eq.4.3.2}) for the found coefficient $f_{0}$ for $^{4}{\rm He}$ on this basis).

\end{enumerate}

\item
The wave functions for the nucleons and the nucleon densities obtained by the DOS model and RMF theory are essentially different.
On example of the $^{4}{\rm He}$ nucleus in the ground state
we observe a clear radial distance between maximums of the function of the DOS model [$h_{1}(r_{\rm max})=8.31769\,{\rm MeV}^{1/2}$ at $r_{\rm max}=1.2024$~fm]
and functions of RMF theory
[$r_{\rm max}\, g(r_{\rm max})=162.29505\,{\rm MeV}^{1/2}$ at $r_{\rm max}=4.64930$~fm, and
$r_{\rm max}\, f(r_{\rm max})=8.73757 \cdot 10^{8}\,{\rm MeV}^{1/2}$ at $r_{\rm max}=4.32866$~fm]
(see Fig.~\ref{fig.5.1}).
We conclude the following.

\begin{itemize}
\item
Such a difference is explained by appearance of the additional factor $|V_{1}(r)|^{1/2}$ in the radial wave function $g(r)$ and
the corresponding factor $|V_{1}|^{-1/2}\, ( - V_{1}^{\prime}/(2V_{1}) + r a_{r}^{-2} )$ in the radial wave function $f(r)$ [see Eqs.~(\ref{eq.5.1.9})].

\item
A probability of nucleons described by the wave function $f(r)$ is essentially larger than the probability given by the function~$g(r)$.
So, for more quick (simple) calculations we can recommend to use only the function $f(r)$ in analysis of quantum properties of this nucleus.

\item
The nucleon densities given by RMF theory contradict to knowledge about distribution of the proton and neutron densities inside the nuclei obtained from experimental study of
scattering of high energetic electrons off nuclei,
spectroscopic information about transitions in muon atoms,
isotopic and isomer shifts in the optical spectra
(see Refs.~\cite{De_Jager.1974.ADNDT,De_Vries.1987.ADNDT,Angeli.2004.ADNDT}).
This indicates that the functions $g$ and $f$ have no sense as the wave functions of quantum physics.
At the same time, the function $h$ which is part of these two functions above, provides us a proper description of quantum properties of nucleons inside the nucleus.

\end{itemize}

\item
We calculate meson function $w^{0}(\mathbf{r})$ (see Eqs.~(\ref{eq.6.2}), (\ref{eq.6.5}) and (\ref{eq.6.8})) which corresponds to the found nucleon density above (see Fig.~\ref{fig.6.1}).
We observe similarity of its radial shape (after re-scaling) with the shape of the nucleon radial density given in Figs.~\ref{fig.5.1}(b) and \ref{fig.5.2}.

\item
The potential $V(r)$ defined in Eq.~(45) in Ref.~\cite{Ring.1996.PPNP} on the basis of the found meson wave function $w_{0}$ above, is connected with
the wave function $\psi_{\varepsilon jlM}(\mathbf{r})$ in Eqs.~(\ref{eq.2.2.2}) and (\ref{eq.2.2.3})
but this wave functions is not exact solution of Dirac Eqs.~(\ref{eq.2.2.1}) with this potential.
On the other side, we found the potential $V_{1}$ which is directly connected with the potential of the DOS model.
If the meson theory describes quantum properties of nucleus well, then a difference between these two potentials should be as small as possible (tend to zero).
In order to analyze this, we introduce new quantum corrections $g_{\rm quant,w}$, $\Delta V(r)$, and $V_{0}$ in Eq.~(\ref{eq.7.1}) characterizing difference between these
two potentials
[we restrict ourselves by the potential $V(r)$ without terms $g_{\rho} \tau_{3} \rho_{3}^{0} (\mathbf{r})$ and $\varepsilon\, A^{0} (\mathbf{r})$
(electromagnetic interactions)].
In order to find these new corrections, we introduce the following requirement:
\emph{Correction $\Delta V(r)$ should be minimal}.
Using this condition, we find the following.

\begin{enumerate}
\item
The potentials $V_{1}(r)$ and $v_{w}(r)$ are very close and are intersected at point $r_{0} = \sqrt{3}\, a_{r}$ (see Fig.~\ref{fig.7.1}).
The difference between these two potentials is very small inside a small radial region at $r \le r_{0}$,
but at $r \ge r_{0}$ it is highly increased at increasing radius $r$.
The quantum correction $\Delta V(r)$ is practically determined only by the potential $V_{1}(r)$ in this region.

\item
We find an unknown new factor $g_{\rm quant, w}$ from condition of coincidence $V_{1}(r_{0}) = v_{w}(r_{0})$ at the radial point of intersection $r_{0}$ for the ground state (here, $r_{0}=\sqrt{3}\,a_{r}$, see Fig.~\ref{fig.4.1}).
We obtain $g_{\rm quant, w} = 2.3864127 \cdot 10^{6}$.
This indicates that the initial potential $V_{1}$ (and meson function $w^{0}$) should be strongly reinforced in order to minimize maximally the additional correction $\Delta V$.

\item
After inclusion of the factor $q_{\rm quantum,w}$ (and without adding $\Delta V(r)$), the modified meson potential $v_{w}(r)$ can be successful in description of the quantum properties (on the level of the shell model) for analyzing nucleus $^{4}{\rm He}$ for energies below the ground state or a little above (for scattering processes).
It is impossible to describe the quantum properties properly for the energies higher than the ground state energy level without inclusion of the quantum corrections $\Delta V(r)$.
This demonstrates an importance of introduction of the quantum corrections to the RMF theory.

\item
Introduction of the correction $\Delta V(r)$ allows to describe completely the quantum properties of the nucleus $^{4}{\rm He}$, as in the DOS model. The shape of the correction $\Delta V(r)$ is presented in Fig.~\ref{fig.7.2}.
One can see that at $r \le r_{0}$ the correction $\Delta V$ is very small, but it is increased highly at increasing $r$ at $r > r_{0}$.
\end{enumerate}
\end{enumerate}
At finishing, we should like to note that the approach proposed in this paper opens a new perspective of next application of methods of quantum mechanics (oriented on high accuracy in detailed description and study of quantum properties in nuclear physic) to the tasks of the RMF theory
(for example, see Ref.~\cite{Maydanyuk.2015.NPA} for details).

\section*{Acknowledgements
\label{sec.acknowledgements}}

S.~P.~M. thanks the Institute of Modern Physics of Chinese Academy of Sciences for its warm hospitality and support.
This work has been supported by
the Major State Basic Research Development Program in China (No. 2015CB856903),
the National Natural Science Foundation of China (Grant Nos. 11575254, 11447105 and 11175215),
the Chinese Academy of Sciences fellowships for researchers from developing countries (No. 2014FFJA0003), and
the Chinese Academy of Sciences President's International Fellowship Initiative (No. 2015PM062).


\appendix
\section{Transition from the radial Dirac equation to one-dimensional differential equation of the second order
(type of equation of Schr\"{o}dinger-like form)
\label{sec.app.1}}

Let us consider equation of Dirac
\begin{equation}
  i\,\hbar \displaystyle\frac{\partial \Psi}{\partial t} = \hat{H} \Psi
\label{eq.app.1.1}
\end{equation}
with Hamiltonian of form:
\begin{equation}
  \hat{H} = \alpha \mathbf{p} + \beta m + V(r),
\label{eq.app.1.2}
\end{equation}
where $V(r)$ is potential energy, dependent on the radial variable $r$, $\alpha$ and $\beta$ are ÷åòûðåõðÿäíûå matrixes of Dirac, $\mathbf{p} = -i\,\partial / \partial \mathbf{r}$ is operator of momentum of the particle, $m$ is mass of the particle.
We choose a system of units where $c=\hbar=1$ (i.~e. we shall omit these constants sometimes).

We shall find a solution for the wave function in form (see Sect.~1.5.1, page~48 in~\cite{Ahiezer.1981}):
\begin{equation}
\begin{array}{lcl}
  \Psi (\mathbf{r},t) = \psi_{\varepsilon jlM}(\mathbf{r}) \cdot e^{-i \varepsilon t / \hbar}, &
  \psi_{\varepsilon jlM}(\mathbf{r}) =
  \left(
  \begin{array}{lcl}
    \varphi_{\varepsilon jlM} (\mathbf{r}) \\
    \chi_{\varepsilon jlM} (\mathbf{r})
  \end{array}
  \right),
\end{array}
\label{eq.app.1.3}
\end{equation}
where
\begin{equation}
\begin{array}{lcl}
  \varphi_{\varepsilon jlM} (\mathbf{r}) =
   g\,(r)\; \Omega_{jlM} (\mathbf{n}), &
  \chi_{\varepsilon jlM} (\mathbf{r}) =
   i\,f\,(r)\; \Omega_{jl^{\prime}M} (\mathbf{n})
\end{array}
\label{eq.app.1.4}
\end{equation}
and $\Omega_{jlM} (\mathbf{n})$ is
\definition{spherical harmonic spinor} (called also as \definition{spherical harmonic spinor function}), $l^{\prime}=2j-l$, $\mathbf{n} = \mathbf{r}/r$, $r=|\mathbf{r}|$, $g\,(r)$ and $f\,(r)$ are radial functions.
According to~\cite{Ahiezer.1981}, $\varepsilon$ is eigenvalue of Hamiltonian (but not energy of the particle), which can be positive and negative. At $\varepsilon>0$, the equation describes particle, at $\varepsilon<0$ --- antiparticle. The energy is defined as $E = |\varepsilon|$, i.~e. it is positive for the particles, and for the antiparticles.
Substituting formulas~(\ref{eq.app.1.3}) and (\ref{eq.app.1.4}) for the wave function into equation~(\ref{eq.app.1.1}), we obtain system of two equations for the radial functions
(see (1.5.4), page~49 in~\cite{Ahiezer.1981}):
\begin{equation}
\begin{array}{lcl}
  \vspace{2mm}
  \displaystyle\frac{d\, (rg(r))}{dr} +
    \displaystyle\frac{\chi}{r}\, (rg(r)) -
    \bigl[\varepsilon + m - V(r)\bigr]\, (rf(r)) = 0, \\
  \displaystyle\frac{d\, (rf(r))}{dr} -
    \displaystyle\frac{\chi}{r}\, (rf(r)) +
    \bigl[\varepsilon - m - V(r)\bigr]\, (rg(r)) = 0,
\end{array}
\label{eq.app.1.5}
\end{equation}
where $\chi = l(l+1) - j(j+1) - 1/4$ (see (1.4.6), page~44 in~\cite{Ahiezer.1981}).
From this system we remove $f$, in order to obtain equation for determination of $g$.
From the first equation we find:
\begin{equation}
\begin{array}{lcl}
  rf(r) & = &
    \displaystyle\frac{1}{\varepsilon + m - V(r)}\,
    \Bigl\{ \displaystyle\frac{d\, (rg(r))}{dr} + \displaystyle\frac{\chi}{r}\, (rg(r)) \Bigr\}
\end{array}
\label{eq.app.1.6}
\end{equation}
and derivative from this function:
\begin{equation}
\begin{array}{lcl}
  \vspace{2mm}
  \displaystyle\frac{d\, (rf(r))}{dr} & = &
  \displaystyle\frac{d}{dr}\,
  \biggl[
    \displaystyle\frac{1}{\varepsilon + m - V(r)}\,
    \Bigl\{ \displaystyle\frac{d\, (rg(r))}{dr} + \displaystyle\frac{\chi}{r}\, (rg(r)) \Bigr\}
  \biggr] = \\
  \vspace{2mm}
  & = &
  \displaystyle\frac{1}{\varepsilon + m - V(r)}\, \displaystyle\frac{d^{2}\, (rg(r))}{dr^{2}} +
  \biggl[
    \displaystyle\frac{1}{(\varepsilon + m - V(r))^{2}}\, \displaystyle\frac{dV(r)}{dr} +
    \displaystyle\frac{1}{\varepsilon + m - V(r)}\, \displaystyle\frac{\chi}{r}
  \biggr]\, \displaystyle\frac{d\, (rg(r))}{dr} + \\
  \vspace{2mm}
  & + &
  \biggl[
    \displaystyle\frac{1}{(\varepsilon + m - V(r))^{2}}\, \displaystyle\frac{\chi}{r}\,
      \displaystyle\frac{dV(r)}{dr} -
    \displaystyle\frac{1}{\varepsilon + m - V(r)}\, \displaystyle\frac{\chi}{r^{2}}
  \biggr] (rg(r)).
\end{array}
\label{eq.app.1.7}
\end{equation}
We substitute these formulas into the second equation of the system~(\ref{eq.app.1.5}), and obtain:
\begin{equation}
\begin{array}{cl}
  \vspace{2mm}
  &
  \displaystyle\frac{d^{2}\, (rg(r))}{dr^{2}} +
  \displaystyle\frac{1}{\varepsilon + m - V(r)}\, \displaystyle\frac{dV(r)}{dr}\,
    \displaystyle\frac{d\, (rg(r))}{dr}\; + \;
  F(r)\, (rg(r)) = 0,
\end{array}
\label{eq.app.1.8}
\end{equation}
where a new function was introduced:
\begin{equation}
  F(r) =
  \displaystyle\frac{1}{\varepsilon + m - V(r)}\, \displaystyle\frac{\chi}{r}\, \displaystyle\frac{dV(r)}{dr} -
  \displaystyle\frac{\chi (1 + \chi)}{r^{2}} +
  \bigl[\varepsilon - V(r)\bigr]^{2} - m^{2}.
\label{eq.app.1.9}
\end{equation}
Now we use change:
\begin{equation}
  r\,g(r) = a(r)\,h(r),
\label{eq.app.1.10}
\end{equation}
from where we find:
\begin{equation}
\begin{array}{cc}
  (r\,g(r))^{\prime} = a^{\prime}(r)\,h(r) + a(r)\,h(r)^{\prime}, &
  (r\,g(r))^{\prime\prime} = a^{\prime\prime}(r)\,h(r) + 2a^{\prime}(r)\,h^{\prime}(r) + a(r)\,h(r)^{\prime\prime}.
\end{array}
\label{eq.app.1.11}
\end{equation}
Then, we shall rewrite equation~(\ref{eq.app.1.8}) in form:
\begin{equation}
  a^{\prime\prime}\,h\; +\;
  2a^{\prime}\,h^{\prime}\; +\; a\,h^{\prime\prime}\; +\;
  \displaystyle\frac{1}{\varepsilon + m - V}\, \displaystyle\frac{dV}{dr}
  \Bigl\{a^{\prime}\,h + a\,h^{\prime}\Bigr\}\; + \;
  F\,a\,h = 0.
\label{eq.app.1.12}
\end{equation}
Let us transform further as
\begin{equation}
  h^{\prime\prime} +
  \Bigl\{\
    2\,\displaystyle\frac{a^{\prime}}{a} +
    \displaystyle\frac{V^{\prime}}{\varepsilon + m - V}\,
  \Bigr\}\, h^{\prime} +
  \Bigl\{\
    \displaystyle\frac{a^{\prime\prime}}{a} +
    \displaystyle\frac{V^{\prime}}{\varepsilon + m - V}\, \displaystyle\frac{a^{\prime}}{a} +
    F
  \Bigr\}\, h = 0.
\label{eq.app.1.13}
\end{equation}
From this equation we formulate a new condition for determination of the unknown function $a(r)$:
%
\begin{equation}
  2\,\displaystyle\frac{a^{\prime}}{a} + \displaystyle\frac{V^{\prime}}{\varepsilon + m - V} = 0.
\label{eq.app.1.14}
\end{equation}
Then, equation~(\ref{eq.app.1.13}) is transformed to
\begin{equation}
\begin{array}{cc}
  h^{\prime\prime} +
  \Bigl\{\
    \displaystyle\frac{a^{\prime\prime}}{a} +
    \displaystyle\frac{V^{\prime}}{\varepsilon + m - V}\, \displaystyle\frac{a^{\prime}}{a} +
    F
  \Bigr\}\, h = 0.
\end{array}
\label{eq.app.1.15}
\end{equation}
We rewrite it in the final form, introducing a new potential function
(and taking eq.~(\ref{eq.app.1.14}) into account):
\begin{equation}
  -\,h^{\prime\prime} + \bar{V}(r)\, h = 0,
\label{eq.app.1.16}
\end{equation}
where
\begin{equation}
\begin{array}{cc}
  \bar{V}(r) =
  -\, \displaystyle\frac{a^{\prime\prime}}{a} +
  \displaystyle\frac{1}{2}\, \Bigl\{\displaystyle\frac{V^{\prime}}{\varepsilon + m - V}\, \Bigr\}^{2} -
  \displaystyle\frac{V^{\prime}}{\varepsilon + m - V(r)}\, \displaystyle\frac{\chi}{r}\, -
  \bigl[\varepsilon - V(r)\bigr]^{2} + m^{2} +
  \displaystyle\frac{\chi (1 + \chi)}{r^{2}}.
\end{array}
\label{eq.app.1.17}
\end{equation}

From Eq.~(\ref{eq.app.1.14}) we find:
\begin{equation}
\begin{array}{l}
  \displaystyle\frac{a^{\prime\prime}}{a} =
  - \displaystyle\frac{V^{\prime\prime}}{2\, (\varepsilon + m - V)} -
  \displaystyle\frac{(V^{\prime})^{2}}{4\, (\varepsilon + m - V)^{2}}.
\end{array}
\label{eq.app.1.18}
\end{equation}
Substituting this formula to eq.~(\ref{eq.app.1.17}), we find:
\begin{equation}
\begin{array}{cc}
  \bar{V}(r) =
  \displaystyle\frac{V^{\prime\prime}}{2\, (\varepsilon + m - V)} +
  \displaystyle\frac{3\,(V^{\prime})^{2}}{4\, (\varepsilon + m - V)^{2}} -
  \displaystyle\frac{V^{\prime}}{\varepsilon + m - V(r)}\, \displaystyle\frac{\chi}{r}\, -
  \bigl[ \varepsilon - V(r) \bigr]^{2} + m^{2} +
  \displaystyle\frac{\chi (1 + \chi)}{r^{2}}.
\end{array}
\label{eq.app.1.19}
\end{equation}

Now we shall obtain the wave function. From eq.~(\ref{eq.app.1.14}) we find a solution for $a$
(we shall consider case of $a>0$)\footnote{
at $\varepsilon + m - V < 0$ we have the same formula:
$ 2\,\displaystyle\frac{a^{\prime}}{a} =
(\ln a^{2})^{\prime} =
- \displaystyle\frac{V^{\prime}}{\varepsilon + m - V} =
\displaystyle\frac{(V -\varepsilon - m)^{\prime}}{V -\varepsilon - m} =
[\ln \,|V - \varepsilon - m|]^{\prime}$.}:
\begin{equation}
  a = n_{0}\; \sqrt{|\varepsilon + m - V|},
\label{eq.app.1.20}
\end{equation}
where $n_{0}$ is arbitrary constant of integration.
We write the final solution for the radial function $g(r)$:
\begin{equation}
  g\,(r) = \displaystyle\frac{n_{0}\; \sqrt{|\varepsilon + m - V(r)|}}{r} \cdot h\,(r).
\label{eq.app.1.21}
\end{equation}

Now we shall find solution for function $f(r)$.
From eq.~(\ref{eq.app.1.6}) with taking into account eq.~(\ref{eq.app.1.21}), we have
\begin{equation}
  f\,(r) =
    \displaystyle\frac{\pm\, n_{0}}{r\: \sqrt{|\varepsilon + m - V(r)|}}\;
    \biggl\{
      \Bigl[
        \displaystyle\frac{-V^{\prime}(r)}{2\: (\varepsilon + m - V(r))} +
        \displaystyle\frac{\chi}{r}\;
      \Bigr]\; h\,(r) +
      h^{\prime}\,(r)
    \biggr\}.
\label{eq.app.1.22}
\end{equation}
Here, upper sign is used at $\varepsilon + m - V(r) > 0$ and bottom sign --- at $\varepsilon + m - V(r) < 0$.
Introducing a function
\begin{equation}
  {\rm sign}\, \bigl[x \bigr] =
  \left\{
  \begin{array}{ll}
    1,  & \mbox{\rm at } x > 0, \\
    -1, & \mbox{\rm at } x < 0, \\
  \end{array}
  \right.
\label{eq.app.1.23}
\end{equation}
we rewrite solution~(\ref{eq.app.1.22}) in form:
\begin{equation}
  f\,(r) =
    {\rm sign}\, \bigl[\varepsilon + m - V(r) \bigr]\;
    \displaystyle\frac{n_{0}}{r\: \sqrt{|\varepsilon + m - V(r)|}}\;
    \biggl\{
      \Bigl[
        \displaystyle\frac{-V^{\prime}(r)}{2\: (\varepsilon + m - V(r))} +
        \displaystyle\frac{\chi}{r}\;
      \Bigr]\; h\,(r) +
      h^{\prime}\,(r)
    \biggr\}.
\label{eq.app.1.24}
\end{equation}

\section{Search of solutions in form of series
\label{sec.app.3}}

In this section we shall analyze possibility of existence of solution of equation~(\ref{eq.3.1.6})
\begin{equation}
\begin{array}{cc}
  4\, \bar{V}(r)  V_{1}^{2} =
  - 2\, V_{1} V_{1}^{\prime\prime} +
  3\,(V_{1}^{\prime})^{2} -
  \displaystyle\frac{4}{r}\, V_{1} V_{1}^{\prime}\, -
  4\, V_{1}^{4} - 8\, V_{1}^{3}\, m
\end{array}
\label{eq.app.3.1}
\end{equation}
in for of series
\begin{equation}
\begin{array}{lcl}
\vspace{1mm}
  V_{1}(r) & = & f_{0} + \displaystyle\sum\limits_{n=1}^{N} f_{n} r^{n}, \\
\vspace{1mm}
  V_{1}^{\prime}(r) & = &
    f_{1} + \displaystyle\sum\limits_{n=1} (n+1) f_{n+1} r^{n}, \\
  V_{1}^{\prime\prime}(r) & = &
    2f_{2} +
    \displaystyle\sum\limits_{n=1} (n+1)(n+2) f_{n+2} r^{n}.
\end{array}
\label{eq.app.3.2}
\end{equation}
We have property
\begin{equation}
  \displaystyle\sum\limits_{n=0} (n+1) f_{n+1} r^{n} \displaystyle\sum\limits_{k=0} (k+1) f_{k+1} r^{k} =
  \displaystyle\sum\limits_{n=0} \Bigl\{\displaystyle\sum\limits_{s=0}^{n} (s+1) (n-s+1) f_{s+1} f_{n-s+1} \Bigr\}\, r^{n}.
\label{eq.app.3.3}
\end{equation}
At first, we calculate powers of potential $V_{1}$:
\begin{equation}
\begin{array}{lcl}
\vspace{1mm}
  V_{1}^{2} & = &
  \displaystyle\sum\limits_{n=0} \displaystyle\sum\limits_{s=0}^{n} f_{s} f_{n-s}\, r^{n}, \\

\vspace{1.5mm}
  V_{1}^{3} & = &
  \displaystyle\sum\limits_{n=0}
  \Bigl\{
    \displaystyle\sum\limits_{s=0}^{n}
    \displaystyle\sum\limits_{t=0}^{s}
      f_{t} f_{s-t}\, f_{n-s}
  \Bigr\}\, r^{n}, \\

  V_{1}^{4} & = &
  \displaystyle\sum\limits_{n=0}
  \Bigl\{
    \displaystyle\sum\limits_{s=0}^{n}
      \displaystyle\sum\limits_{p=0}^{s}
      \displaystyle\sum\limits_{t=0}^{p}
        f_{t}\, f_{p-t}\, f_{s-p}\, f_{n-s}
  \Bigr\}\,
  r^{n}.
\end{array}
\label{eq.app.3.4}
\end{equation}
The first, second and third terms in the right part of Eq.~(\ref{eq.app.3.1}) are
\begin{equation}
\begin{array}{lcl}
  V_{1} V_{1}^{\prime\prime} & = &
  \displaystyle\sum\limits_{n=0}
  \Bigl\{
    \displaystyle\sum\limits_{s=0}^{n}
    (s+1)(s+2)\, f_{s+2}\, f_{n-s}
  \Bigr\}\, r^{n}, \\

  (V_{1}^{\prime})^{2} & = &
  \displaystyle\sum\limits_{n=0}
  \Bigl\{
    \displaystyle\sum\limits_{s=0}^{n}
      (s+1) f_{s+1}\:
      (n-s+1) f_{n-s+1}
  \Bigr\} \, r^{n}, \\

  \displaystyle\frac{1}{r}\, V_{1} V_{1}^{\prime} & = &
  f_{0} f_{1}\, \displaystyle\frac{1}{r} +
  \displaystyle\sum\limits_{n=0}
  \Bigl\{
    \displaystyle\sum\limits_{s=0}^{n+1}
      (s+1) f_{s+1}\, f_{n-s+1}
  \Bigr\}\; r^{n}.
\end{array}
\label{eq.app.3.5}
\end{equation}
We find summation of terms in the right part of Eq.~(\ref{eq.app.3.1}):
\begin{equation}
\begin{array}{lcl}
\vspace{3mm}
  & &
  - 2\, V_{1} V_{1}^{\prime\prime} +
  3\,(V_{1}^{\prime})^{2} -
  \displaystyle\frac{4}{r}\, V_{1} V_{1}^{\prime}\, -
  4\, V_{1}^{4} - 8\, V_{1}^{3}\, m\; = \\

\vspace{0.5mm}
  & = &
  - 4\, f_{0} f_{1}\, \displaystyle\frac{1}{r} +
  \displaystyle\sum\limits_{n=0}
  \biggl\{
    \displaystyle\sum\limits_{s=0}^{n}
    \Bigl[
      - 2\, (s+1)(s+2)\: f_{s+2}\, f_{n-s} +
      3\, (s+1) (n-s+1)\: f_{s+1}\, f_{n-s+1}\; - \\
  & & -\;
      4\, (s+1) f_{s+1}\, f_{n-s+1} -
      4 \displaystyle\sum\limits_{p=0}^{s}
        \displaystyle\sum\limits_{t=0}^{p}
          f_{t}\, f_{p-t}\, f_{s-p}\, f_{n-s} -
      8m \displaystyle\sum\limits_{t=0}^{s} f_{t} f_{s-t}\, f_{n-s}
    \Bigr]\, -\,
    4\, (n+2)\, f_{0} f_{n+2}
  \biggl\}\, r^{n}.
\end{array}
\label{eq.app.3.6}
\end{equation}
We simplify this expression further. We have
\begin{equation}
\begin{array}{lcl}
  & &
  \displaystyle\sum\limits_{s=0}^{n}
    - 2\, (s+1)(s+2)\: f_{s+2}\, f_{n-s} +
  \displaystyle\sum\limits_{s=0}^{n}
    3\, (s+1) (n-s+1)\: f_{s+1}\, f_{n-s+1}\; = \\
  & = &
  - 4\, f_{2}\, f_{n} +
  2n\, (n+1)\, f_{1}\, f_{n+1} +
  \displaystyle\sum\limits_{s=0}^{n}
    (s+1)\, (3n - 5s + 3 )\, f_{s+1}\, f_{n-s+1}
\end{array}
\label{eq.app.3.7}
\end{equation}
and obtain
\begin{equation}
\begin{array}{lcl}
  & &
  - 2\, V_{1} V_{1}^{\prime\prime} +
  3\,(V_{1}^{\prime})^{2} -
  \displaystyle\frac{4}{r}\, V_{1} V_{1}^{\prime}\, -
  4\, V_{1}^{4} - 8\, V_{1}^{3}\, m\; = \\

  & = &
  - 4\, f_{0} f_{1}\, \displaystyle\frac{1}{r} +
  \displaystyle\sum\limits_{n=0}
  \biggl\{
    \displaystyle\sum\limits_{s=0}^{n}
    \Bigl[
      (s+1)\, (3n - 5s - 1)\, f_{s+1}\, f_{n-s+1} -
      4 \displaystyle\sum\limits_{p=0}^{s}
        \displaystyle\sum\limits_{t=0}^{p}
          f_{t}\, f_{p-t}\, f_{s-p}\, f_{n-s} -
      8m \displaystyle\sum\limits_{t=0}^{s} f_{t} f_{s-t}\, f_{n-s}
    \Bigr]\; - \\

  & & -\;
    4\, f_{2}\, f_{n} +
    2n\, (n+1)\, f_{1}\, f_{n+1} -\,
    4\, (n+2)\, f_{0} f_{n+2}
  \biggl\}\, r^{n}.
\end{array}
\label{eq.app.3.8}
\end{equation}

Now we analyze the potential in a form
\begin{equation}
  \bar{V} (r) = a_{0} + a_{1} r + a_{2} r^{2},
\label{eq.app.3.9}
\end{equation}
where $a_{0}$, $a_{1}$ and $a_{3}$ are coefficients defined from the microscopic model. We obtain
(we assume $f_{-1} = f_{-2} = 0$):
\begin{equation}
\begin{array}{lcl}
  4\, \bar{V}\, V_{1}^{2} & = &
  4\, a_{0}\, f_{0}^{2} +
  4\, f_{0} (2a_{0} f_{1} + a_{1} f_{0})\, r\;\; +\;\;
  4\, \displaystyle\sum\limits_{n=2} \displaystyle\sum\limits_{s=0}^{n}
    f_{s}\: \bigl[ a_{0} f_{n-s} + a_{1} f_{n-s-1} + a_{2} f_{n-s-2} \bigr]\: r^{n}.
\end{array}
\label{eq.app.3.10}
\end{equation}
Taking into account that this equation has no terms $r^{-n}$ with negative powers, from Eq.~(\ref{eq.app.3.8}) we obtain:
\begin{equation}
  f_{1} = 0.
\label{eq.app.3.11}
\end{equation}
Taking this into account, we simplify Eq.~(\ref{eq.app.3.8}) as
\begin{equation}
\begin{array}{lcl}
\vspace{1mm}
  & &
  - 2\, V_{1} V_{1}^{\prime\prime} +
  3\,(V_{1}^{\prime})^{2} -
  \displaystyle\frac{4}{r}\, V_{1} V_{1}^{\prime}\, -
  4\, V_{1}^{4} - 8\, V_{1}^{3}\, m\; = \\

  & = &
  \displaystyle\sum\limits_{n=0}
  \biggl\{
    \displaystyle\sum\limits_{s=1}^{n}
    \Bigl[
      (s+1)\, (3n - 5s - 1)\, f_{s+1}\, f_{n-s+1} -
      4 \displaystyle\sum\limits_{p=0}^{s}
        \displaystyle\sum\limits_{t=0}^{p}
          f_{t}\, f_{p-t}\, f_{s-p}\, f_{n-s} -
      8m \displaystyle\sum\limits_{t=0}^{s} f_{t} f_{s-t}\, f_{n-s}
    \Bigr]\; - \\

  & & -\;
    4\, f_{2}\, f_{n} -
    4\, (n+2)\, f_{0} f_{n+2}
  \biggl\}\, r^{n}.
\end{array}
\label{eq.app.3.12}
\end{equation}
Expression~(\ref{eq.app.3.10}) is transformed into the following
\begin{equation}
\begin{array}{lcl}
  4\, \bar{V}\, V_{1}^{2} & = &
  4\, a_{0}\, f_{0}^{2} +
  4\, a_{1} f_{0}^{2}\, r\;\; +\;\;
  4\, \displaystyle\sum\limits_{n=2} \displaystyle\sum\limits_{s=0}^{n}
    f_{s}\: \bigl[ a_{0} f_{n-s} + a_{1} f_{n-s-1} + a_{2} f_{n-s-2} \bigr]\: r^{n}.
\end{array}
\label{eq.app.3.13}
\end{equation}
Taking into account the initial equation (\ref{eq.app.3.7}), we obtain equation for determination the unknown amplitudes $f_{n}$
\begin{equation}
\begin{array}{lcl}
\vspace{1mm}
  & & 4\, a_{0}\, f_{0}^{2} +
  4\, a_{1} f_{0}^{2}\, r\;\; +\;\;
  4\, \displaystyle\sum\limits_{n=2} \displaystyle\sum\limits_{s=0}^{n}
    f_{s}\: \bigl[ a_{0} f_{n-s} + a_{1} f_{n-s-1} + a_{2} f_{n-s-2} \bigr]\: r^{n} = \\

\vspace{-0.5mm}
  & = &
  \Bigl\{ - 4 f_{0}^{4} - 8m f_{0}^{3} - 12 f_{0} f_{2} \Bigl\}\; + \\
\vspace{-0.5mm}
  & + &
  \biggl\{
    - \displaystyle\sum\limits_{s=0}^{1}
    \Bigl[
     4 \displaystyle\sum\limits_{p=0}^{s}
        \displaystyle\sum\limits_{t=0}^{p}
          f_{t} f_{p-t} f_{s-p} f_{1-s} +
      8m \displaystyle\sum\limits_{t=0}^{s} f_{t} f_{s-t} f_{1-s}
    \Bigr]\, -\,
    12\, f_{0} f_{3}
  \biggl\}\, r\; + \\

\vspace{-1.0mm}
  & + &
  \displaystyle\sum\limits_{n=2}
  \biggl\{
    \displaystyle\sum\limits_{s=0}^{n}
    \Bigl[
      (s+1)\, (3n - 5s - 1)\, f_{s+1}\, f_{n-s+1} -
      4 \displaystyle\sum\limits_{p=0}^{s}
        \displaystyle\sum\limits_{t=0}^{p}
          f_{t}\, f_{p-t}\, f_{s-p}\, f_{n-s} -
      8m \displaystyle\sum\limits_{t=0}^{s} f_{t} f_{s-t}\, f_{n-s}
    \Bigr]\; - \\
  & & -\;
    4\, f_{2}\, f_{n} - 4\, (n+2)\, f_{0} f_{n+2}
  \biggl\}\, r^{n}.
\end{array}
\label{eq.app.3.14}
\end{equation}
From it, we obtain solutions for the coefficients $f_{n}$ at different powers of $n$ ($n \ge 2$)
\begin{equation}
\begin{array}{lcl}
\vspace{1.5mm}
  f_{1} & = & 0, \\

\vspace{1.5mm}
  f_{2} & = & - \displaystyle\frac{f_{0}}{3}\, \bigl[ a_{0} + 2m f_{0} + f_{0}^{2} \bigr], \\

\vspace{1.5mm}
  f_{3} & = & -\, \displaystyle\frac{a_{1} f_{0}}{3}, \\

\vspace{0.5mm}
  f_{n+2} & = &
  -\, \displaystyle\frac{f_{2} f_{n}}{(n+2)\, f_{0}}\: +\:
  \displaystyle\frac{1}{(n+2)\, f_{0}}\,
  \displaystyle\sum\limits_{s=0}^{n}
  \Bigl[
    \displaystyle\frac{(s+1)\, (3n - 5s - 1)}{4}\, f_{s+1}\, f_{n-s+1}\; - \\
  & - &
    \displaystyle\sum\limits_{p=0}^{s}
    \displaystyle\sum\limits_{t=0}^{p}
      f_{t}\, f_{p-t}\, f_{s-p}\, f_{n-s} -
    2m \displaystyle\sum\limits_{t=0}^{s} f_{t} f_{s-t}\, f_{n-s} -
    f_{s}\, \bigl( a_{0} f_{n-s} + a_{1} f_{n-s-1} + a_{2} f_{n-s-2} \bigr)
  \Bigr].
\end{array}
\label{eq.app.3.18}
\end{equation}

\section{Calculations of coefficients $c_{n}$
\label{sec.app.4}}

In this Section we shall find the unknown coefficients in Eq.~(\ref{eq.5.2.8}).
We combine Eqs.~(\ref{eq.5.2.7}) and (\ref{eq.5.2.8}) as
\begin{equation}
\begin{array}{lllll}
  \rho_{v} (\mathbf{r}) & = &
  n_{0}^{2}\,
  \biggl\{
    \Bigl|f_{0} + \displaystyle\sum\limits_{n=1} f_{n} r^{n} \Bigr| +
    \displaystyle\frac{1}{\bigl| f_{0} + \displaystyle\sum\limits_{n=1} f_{n} r^{n} \bigr|}\;
    \biggl|
      - \displaystyle\frac{f_{1} + \displaystyle\sum\limits_{n=1} (n+1) f_{n+1} r^{n}}
        {2\, (f_{0} + \displaystyle\sum\limits_{n=1} f_{n} r^{n})} +
      \displaystyle\frac{r}{a_{r}^{2}}
    \biggr|^{2}
  \biggr\}\,
  \displaystyle\frac{h_{1}^{2}(r)}{r^{2}} =
  n_{0}^{2}\,
  \displaystyle\frac{h_{1}^{2}\,(r)}{r^{2}}\,
  \displaystyle\sum\limits_{n=0} c_{n} r^{n}
\end{array}
\label{eq.app.4.1}
\end{equation}
and obtain
\begin{equation}
\begin{array}{lllll}
  \displaystyle\sum\limits_{n=0} c_{n} r^{n} =
    \Bigl|f_{0} + \displaystyle\sum\limits_{n=1} f_{n} r^{n} \Bigr| +
    \displaystyle\frac{1}{\bigl| f_{0} + \displaystyle\sum\limits_{n=1} f_{n} r^{n} \bigr|}\;
    \biggl|
      - \displaystyle\frac{f_{1} + \displaystyle\sum\limits_{n=1} (n+1) f_{n+1} r^{n}}
        {2\, (f_{0} + \displaystyle\sum\limits_{n=1} f_{n} r^{n})} +
      \displaystyle\frac{r}{a_{r}^{2}}
    \biggr|^{2}.
\end{array}
\label{eq.app.4.2}
\end{equation}
Now we consider a case when the potential $V_{1}$ is positive.
We have
\begin{equation}
\begin{array}{lllll}
  \displaystyle\sum\limits_{n=0} c_{n} r^{n} \Bigl( \displaystyle\sum\limits_{n=0} f_{n} r^{n} \Bigr)^{3} =
  \Bigl(\displaystyle\sum\limits_{n=0} f_{n} r^{n} \Bigr)^{4} +
  \displaystyle\frac{1}{4}\, \Bigl( \displaystyle\sum\limits_{n=0} (n+1) f_{n+1} r^{n} \Bigr)^{2} -
  \displaystyle\frac{r}{a_{r}^{2}}\, \displaystyle\sum\limits_{n=0} (n+1) f_{n+1} r^{n} \displaystyle\sum\limits_{n=0} f_{n} r^{n} +
  \displaystyle\frac{r^{2}}{a_{r}^{4}}\, \Bigl( \displaystyle\sum\limits_{n=0} f_{n} r^{n} \Bigr)^{2}.
\end{array}
\label{eq.app.4.4}
\end{equation}
We use solutions (\ref{eq.app.3.4}) for summations of series and find
\begin{equation}
\begin{array}{lllll}
  \displaystyle\sum\limits_{n=0} c_{n} r^{n} \Bigl( \displaystyle\sum\limits_{n=0} f_{n} r^{n} \Bigr)^{3} =
  \displaystyle\sum\limits_{n=0}
  \Bigl\{
    \displaystyle\sum\limits_{s=0}^{n}
      \displaystyle\sum\limits_{p=0}^{s}
      \displaystyle\sum\limits_{t=0}^{p}
        f_{t}\, f_{p-t}\, f_{s-p}\, c_{n-s}
  \Bigr\}\, r^{n}, \\

  \Bigl(\displaystyle\sum\limits_{n=0} f_{n} r^{n} \Bigr)^{4} =
  \displaystyle\sum\limits_{n=0}
  \Bigl\{
    \displaystyle\sum\limits_{s=0}^{n}
      \displaystyle\sum\limits_{p=0}^{s}
      \displaystyle\sum\limits_{t=0}^{p}
        f_{t}\, f_{p-t}\, f_{s-p}\, f_{n-s}
  \Bigr\}\, r^{n}, \\

  \Bigl( \displaystyle\sum\limits_{n=0} (n+1) f_{n+1} r^{n} \Bigr)^{2} =
  \displaystyle\sum\limits_{n=0} \displaystyle\sum\limits_{s=0}^{n}
    (s+1) (n-s+1)\, f_{s+1} f_{n-s+1}\, r^{n}, \\

  \displaystyle\frac{r}{a_{r}^{2}} \displaystyle\sum\limits_{n=0} (n+1) f_{n+1} r^{n} \displaystyle\sum\limits_{n=0} f_{n} r^{n} =
  \displaystyle\frac{1}{a_{r}^{2}}
    \displaystyle\sum\limits_{n=1} \displaystyle\sum\limits_{s=0}^{n-1}
    (s+1)\, f_{s+1} f_{n-s-1}\, r^{n}, \\

  \displaystyle\frac{r^{2}}{a_{r}^{4}}\, \Bigl( \displaystyle\sum\limits_{n=0} f_{n} r^{n} \Bigr)^{2} =
  \displaystyle\frac{1}{a_{r}^{4}}\,
    \displaystyle\sum\limits_{n=2} \displaystyle\sum\limits_{s=0}^{n-2} f_{s} f_{n-s-2}\, r^{n}.
\end{array}
\label{eq.app.4.6}
\end{equation}
We transform Eq.~(\ref{eq.app.4.4}) as
\begin{equation}
\begin{array}{lllll}
  \displaystyle\sum\limits_{n=0}
  \Bigl\{
    \displaystyle\sum\limits_{s=0}^{n}
      \displaystyle\sum\limits_{p=0}^{s}
      \displaystyle\sum\limits_{t=0}^{p}
        f_{t}\, f_{p-t}\, f_{s-p}\, c_{n-s}
  \Bigr\}\, r^{n} =

  \displaystyle\sum\limits_{n=0}
  \Bigl\{
    \displaystyle\sum\limits_{s=0}^{n}
      \displaystyle\sum\limits_{p=0}^{s}
      \displaystyle\sum\limits_{t=0}^{p}
        f_{t}\, f_{p-t}\, f_{s-p}\, f_{n-s}
  \Bigr\}\, r^{n} +

  \displaystyle\frac{1}{4}\,
    \displaystyle\sum\limits_{n=0} \displaystyle\sum\limits_{s=0}^{n}
      (s+1) (n-s+1)\, f_{s+1} f_{n-s+1}\, r^{n} - \\

  - \displaystyle\frac{1}{a_{r}^{2}}
    \displaystyle\sum\limits_{n=1} \displaystyle\sum\limits_{s=0}^{n-1}
    (s+1)\, f_{s+1} f_{n-s-1}\, r^{n} +

  \displaystyle\frac{1}{a_{r}^{4}}\,
    \displaystyle\sum\limits_{n=2} \displaystyle\sum\limits_{s=0}^{n-2} f_{s} f_{n-s-2}\, r^{n}.
\end{array}
\label{eq.app.4.7}
\end{equation}
We solve this equation for different powers of variable $r$ and obtain
\begin{equation}
\begin{array}{lllll}
  c_{0} = f_{0} + \displaystyle\frac{f_{1}^{2}}{4f_{0}^{3}}, \\
  c_{1} = A_{1} f_{0}^{-3} - 3 f_{0}^{-1} f_{1}\, c_{0}, \\
  c_{n} =
  A_{2} f_{0}^{-3} -
  f_{0}^{-3} \displaystyle\sum\limits_{s=1}^{n}
    \displaystyle\sum\limits_{p=0}^{s}
    \displaystyle\sum\limits_{t=0}^{p} f_{t}\, f_{p-t}\, f_{s-p}\, c_{n-s},
\end{array}
\label{eq.app.4.15}
\end{equation}
where
\begin{equation}
\begin{array}{lllll}
  A_{1} & = &
  4\, f_{0}^{3} f_{1} +
  f_{1} f_{2} -
  \displaystyle\frac{1}{a_{r}^{2}} f_{0} f_{1}, \\

  A_{2} & = &
  \displaystyle\sum\limits_{s=0}^{n}
    \displaystyle\sum\limits_{p=0}^{s}
    \displaystyle\sum\limits_{t=0}^{p}
      f_{t}\, f_{p-t}\, f_{s-p}\, f_{n-s} +
  \displaystyle\frac{1}{4}\,
    \displaystyle\sum\limits_{s=0}^{n}
    (s+1) (n-s+1)\, f_{s+1} f_{n-s+1} - \\
  & - &
  \displaystyle\frac{1}{a_{r}^{2}}
    \displaystyle\sum\limits_{s=0}^{n-1} (s+1)\, f_{s+1} f_{n-s-1} +
  \displaystyle\frac{1}{a_{r}^{4}}\, \displaystyle\sum\limits_{s=0}^{n-2} f_{s} f_{n-s-2}.
\end{array}
\label{eq.app.4.16}
\end{equation}

\section{Calculations of coefficients $c_{n,w}$ for the meson function $\chi_{w}(r)$
\label{sec.app.5}}

In this Section we shall find the unknown coefficients $c_{n,w}$ for the meson function $\chi_{w}(r)$ defined in Eq.~(\ref{eq.6.5}) as
\begin{equation}
\begin{array}{lllll}
 \chi_{w} (r) & = &
  n_{0}^{2}\,
  \displaystyle\frac{h^{2}(r)}{r}\,
  \displaystyle\sum\limits_{n=0} c_{w,n} r^{n}.
\end{array}
\label{eq.app.5.1}
\end{equation}
We calculate derivatives of this function
\begin{equation}
\begin{array}{lllll}
\vspace{2mm}
  \chi_{w}^{\prime} (r) & = &
  n_{0}^{2}\,
  \displaystyle\frac{h^{2}(r)}{r}\,
  \biggl\{
    c_{w,0}\, r^{-1} +
    2\, c_{w,1} +
    \displaystyle\sum\limits_{n=1} [(n+2)\, c_{w,n+1} - 2a_{r}^{-2}c_{w,n-1} ]\, r^{n}
  \biggr\}, \\

\vspace{-0.2mm}
  \chi_{w}^{\prime\prime} (r) & = &
  n_{0}^{2}\,
  \displaystyle\frac{h^{2}(r)}{r}\;
  \biggl\{
    \displaystyle\frac{2\, c_{w,1}}{r} +
    6\, \bigl[c_{w,2} - a_{r}^{-2}c_{w,0} \bigr] +
    \bigl[12\, c_{w,3} - 10a_{r}^{-2}c_{w,1} \bigr]\, r\; - \\
  & + &
    \displaystyle\sum\limits_{n=2}
      \Bigl[
         4a_{r}^{-4}c_{w,n-2} - 2a_{r}^{-2} (2n+3)\, c_{w,n} + (n+2)\,(n+3)\, c_{w,n+2}
      \Bigr]\, r^{n}
  \biggr\}.
\end{array}
\label{eq.app.5.2}
\end{equation}
Now we consider equation (\ref{eq.6.4}), having form:
\begin{equation}
\begin{array}{lllll}
\vspace{0.5mm}
  - \chi_{w}^{\prime\prime} (r) + m_{w} \chi_{w} & = &
  - n_{0}^{2}\,
  \displaystyle\frac{h^{2}(r)}{r}\;
  \biggl\{
    \displaystyle\frac{2\, c_{w,1}}{r} +
    6\, \bigl[c_{w,2} - a_{r}^{-2}c_{w,0} \bigr] +
    \bigl[12\, c_{w,3} - 10a_{r}^{-2}c_{w,1} \bigr]\, r\; - \\
  & + &
    \displaystyle\sum\limits_{n=2}
      \Bigl[
         4a_{r}^{-4}c_{w,n-2} - 2a_{r}^{-2} (2n+3)\, c_{w,n} + (n+2)\,(n+3)\, c_{w,n+2}
      \Bigr]\, r^{n}
  \biggr\}\; +\;

  m_{w}\, n_{0}^{2}
    \displaystyle\frac{h^{2}(r)}{r}
    \displaystyle\sum\limits_{n=0} c_{w,n} r^{n}\; = \\

  & = &
  g_{w}\, \rho_{v} (\mathbf{r})\; r =
  g_{w}\; n_{0}^{2}\,
    \displaystyle\frac{h^{2}\,(r)}{r}\,
    \displaystyle\sum\limits_{n=0} c_{n} r^{n}
\end{array}
\label{eq.app.5.3}
\end{equation}
which is transformed to
\begin{equation}
\begin{array}{lllll}
\vspace{-0.5mm}
  & - &
  \biggl\{
    \displaystyle\frac{2\, c_{w,1}}{r} +
    6\, \bigl[c_{w,2} - a_{r}^{-2}c_{w,0} \bigr] +
    \bigl[12\, c_{w,3} - 10a_{r}^{-2}c_{w,1} \bigr]\, r\; - \\
  & + &
    \displaystyle\sum\limits_{n=2}
      \Bigl[
         4a_{r}^{-4}c_{w,n-2} - 2a_{r}^{-2} (2n+3)\, c_{w,n} + (n+2)\,(n+3)\, c_{w,n+2}
      \Bigr]\, r^{n}
  \biggr\}\; +\;
  \displaystyle\sum\limits_{n=0} (m_{w}c_{w,n} - g_{w} c_{n})\, r^{n}\; = 0.
\end{array}
\label{eq.app.5.4}
\end{equation}
From this equation we obtain ($n \ge 2$)
\begin{equation}
\begin{array}{lllll}
\vspace{2mm}
  c_{w,1} = 0, \\

\vspace{2mm}
  c_{w,2} = \Bigl( a_{r}^{-2} + \displaystyle\frac{m_{w}}{6} \Bigr)\,c_{w,0} - \displaystyle\frac{g_{w}}{6}\, c_{0}, \\

\vspace{2mm}
  c_{w,3} = - \displaystyle\frac{g_{w}}{12}\, c_{1}, \\

  c_{w,n+2} =
    - \displaystyle\frac{4a_{r}^{-4}}{(n+2)\,(n+3)}\, c_{w,n-2} +
    \displaystyle\frac{2a_{r}^{-2} (2n+3) + m_{w}}{(n+2)\,(n+3)}\, c_{w,n} -
    \displaystyle\frac{g_{w}}{(n+2)\,(n+3)}\, c_{n}.
\end{array}
\label{eq.app.5.9}
\end{equation}


\end{document}